\pdfoutput=1
\documentclass[12pt,a4paper]{article}

\usepackage{ifthen} 
\newboolean{pdflatex}
\setboolean{pdflatex}{true} 

\newboolean{articletitles}
\setboolean{articletitles}{true} 

\newboolean{uprightparticles}
\setboolean{uprightparticles}{false} 

\def\paperauthors{LHCb collaboration} 
\def\paperasciititle{Study of the rare decay J/psi->4mu} 
\def\papertitle{Study of the rare decay $J\mskip -3mu/\mskip -2mu\psi \to \mu^+\mu^-\mu^+\mu^-$} 
\def\paperkeywords{{High Energy Physics}, {LHCb}} 
\def\papercopyright{\the\year\ CERN for the benefit of the LHCb collaboration} 
\def\paperlicence{CC BY 4.0 licence}
\def\paperlicenceurl{https://creativecommons.org/licenses/by/4.0/}

\usepackage[top=1in, bottom=1.25in, left=1in, right=1in]{geometry}

\usepackage{microtype}
\usepackage{lineno}  
\usepackage{xspace} 
\usepackage{caption} 

\usepackage{graphicx}  
\usepackage{color}
\usepackage{colortbl}
\graphicspath{{./figs/}} 

\usepackage{amsmath} 
\usepackage{amssymb}
\usepackage{amsfonts}
\usepackage{upgreek} 
\usepackage{subcaption}
\newcommand*\patchAmsMathEnvironmentForLineno[1]{%
\expandafter\let\csname old#1\expandafter\endcsname\csname #1\endcsname
\expandafter\let\csname oldend#1\expandafter\endcsname\csname
end#1\endcsname
 \renewenvironment{#1}%
   {\linenomath\csname old#1\endcsname}%
   {\csname oldend#1\endcsname\endlinenomath}%
}
\newcommand*\patchBothAmsMathEnvironmentsForLineno[1]{%
  \patchAmsMathEnvironmentForLineno{#1}%
  \patchAmsMathEnvironmentForLineno{#1*}%
}
\AtBeginDocument{%
\patchBothAmsMathEnvironmentsForLineno{equation}%
\patchBothAmsMathEnvironmentsForLineno{align}%
\patchBothAmsMathEnvironmentsForLineno{flalign}%
\patchBothAmsMathEnvironmentsForLineno{alignat}%
\patchBothAmsMathEnvironmentsForLineno{gather}%
\patchBothAmsMathEnvironmentsForLineno{multline}%
\patchBothAmsMathEnvironmentsForLineno{eqnarray}%
}

\usepackage{hyperxmp}

\usepackage[pdftex,
            pdfauthor={\paperauthors},
            pdftitle={\paperasciititle},
            pdfkeywords={\paperkeywords},
            pdfcopyright={Copyright (C) \papercopyright},
            pdflicenseurl={\paperlicenceurl}]{hyperref}

\usepackage[colorinlistoftodos,textsize=scriptsize]{todonotes}

\usepackage[bottom,flushmargin,hang,multiple]{footmisc}

\usepackage[all]{hypcap} 

\usepackage{xspace} 
\usepackage{upgreek}

\def\lhcb   {\mbox{LHCb}\xspace}

\def\cms    {\mbox{CMS}\xspace}

\def\besiii {\mbox{BESIII}\xspace}

\def\MagUp {\mbox{\em Mag\kern -0.05em Up}\xspace}

\ifthenelse{\boolean{uprightparticles}}%
{

 \def\Peta        {\ensuremath{\upeta}\xspace}

 \def\Pmu         {\ensuremath{\upmu}\xspace}

 \def\Ppi         {\ensuremath{\uppi}\xspace}

 \def\Ppsi        {\ensuremath{\uppsi}\xspace}

 \def\PDelta      {\ensuremath{\Delta}\xspace}                 
 \def\PXi         {\ensuremath{\Xi}\xspace}                 
 \def\PLambda     {\ensuremath{\Lambda}\xspace}                 
 \def\PSigma      {\ensuremath{\Sigma}\xspace}                 
 \def\POmega      {\ensuremath{\Omega}\xspace}                 
 \def\PUpsilon    {\ensuremath{\Upsilon}\xspace}
 \let\oldPi\Pi
 \def\PPi         {\ensuremath{\oldPi}\xspace}

 \def\PB      {\ensuremath{\mathrm{B}}\xspace}                 
                  
 \def\PD      {\ensuremath{\mathrm{D}}\xspace}

 \def\PJ      {\ensuremath{\mathrm{J}}\xspace}                 
 \def\PK      {\ensuremath{\mathrm{K}}\xspace}

 \def\Pb      {\ensuremath{\mathrm{b}}\xspace}                 
 \def\Pc      {\ensuremath{\mathrm{c}}\xspace}                 
                  
 \def\Pe      {\ensuremath{\mathrm{e}}\xspace}

 \def\Pi      {\ensuremath{\mathrm{i}}\xspace}

 \def\Ps      {\ensuremath{\mathrm{s}}\xspace}

 \def\thebaroffset{0.0em}
}
{

 \def\Peta        {\ensuremath{\eta}\xspace}

 \def\Pmu         {\ensuremath{\mu}\xspace}

 \def\Ppi         {\ensuremath{\pi}\xspace}

 \def\Ppsi        {\ensuremath{\psi}\xspace}                 
                  
 \mathchardef\PDelta="7101
 \mathchardef\PXi="7104
 \mathchardef\PLambda="7103
 \mathchardef\PSigma="7106
 \mathchardef\POmega="710A
 \mathchardef\PUpsilon="7107
 \mathchardef\PPi="7105
                  
 \def\PB      {\ensuremath{B}\xspace}                 
                  
 \def\PD      {\ensuremath{D}\xspace}

 \def\PJ      {\ensuremath{J}\xspace}                 
 \def\PK      {\ensuremath{K}\xspace}

 \def\Pb      {\ensuremath{b}\xspace}                 
 \def\Pc      {\ensuremath{c}\xspace}                 
                  
 \def\Pe      {\ensuremath{e}\xspace}

 \def\Pi      {\ensuremath{i}\xspace}

 \def\Ps      {\ensuremath{s}\xspace}

 \def\thebaroffset{0.18em}
}
\newcommand{\offsetoverline}[2][\thebaroffset]{\kern #1\overline{\kern -#1 #2}}%

\makeatletter
\ifcase \@ptsize \relax
  \newcommand{\miniscule}{\@setfontsize\miniscule{4}{5}}
\or
  \newcommand{\miniscule}{\@setfontsize\miniscule{5}{6}}
\or
  \newcommand{\miniscule}{\@setfontsize\miniscule{5}{6}}
\fi
\makeatother

\DeclareRobustCommand{\optbar}[1]{\shortstack{{\miniscule (\rule[.5ex]{1.25em}{.18mm})}
  \\ [-.7ex] $#1$}}

\def\epem       {{\ensuremath{\Pe^+\Pe^-}}\xspace}

\def\mup        {{\ensuremath{\Pmu^+}}\xspace}
\def\mun        {{\ensuremath{\Pmu^-}}\xspace} 

\def\mumu       {{\ensuremath{\Pmu^+\Pmu^-}}\xspace}

\def\squark    {{\ensuremath{\Ps}}\xspace}

\def\cquark    {{\ensuremath{\Pc}}\xspace}
\def\cquarkbar {{\ensuremath{\overline \cquark}}\xspace}
\def\ccbar     {{\ensuremath{\cquark\cquarkbar}}\xspace}
\def\bquark    {{\ensuremath{\Pb}}\xspace}
\def\bquarkbar {{\ensuremath{\overline \bquark}}\xspace}
\def\bbbar     {{\ensuremath{\bquark\bquarkbar}}\xspace}

\def\pion   {{\ensuremath{\Ppi}}\xspace}
\def\piz    {{\ensuremath{\pion^0}}\xspace}
\def\pip    {{\ensuremath{\pion^+}}\xspace}
\def\pim    {{\ensuremath{\pion^-}}\xspace}

\def\KorKbar {\kern \thebaroffset\optbar{\kern -\thebaroffset \PK}{}\xspace}

\newcommand{\etaz}{\ensuremath{\Peta}\xspace}
\newcommand{\etapr}{\ensuremath{\Peta^{\prime}}\xspace}

\def\D       {{\ensuremath{\PD}}\xspace}

\def\DorDbar {\kern \thebaroffset\optbar{\kern -\thebaroffset \PD}\xspace}

\def\Dp      {{\ensuremath{\D^+}}\xspace}
\def\Dm      {{\ensuremath{\D^-}}\xspace}

\def\DpDm    {\ensuremath{\Dp {\kern -0.16em \Dm}}\xspace}

\def\B       {{\ensuremath{\PB}}\xspace}

\def\BorBbar {\kern \thebaroffset\optbar{\kern -\thebaroffset \PB}\xspace}

\def\Bd      {{\ensuremath{\B^0}}\xspace}

\def\BdorBdbar {\kern \thebaroffset\optbar{\kern -\thebaroffset \Bd}\xspace}

\def\Bs      {{\ensuremath{\B^0_\squark}}\xspace}

\def\BsorBsbar {\kern \thebaroffset\optbar{\kern -\thebaroffset \Bs}\xspace}

\def\Bds     {{\ensuremath{\B_{(\squark)}^0}}\xspace}

\def\jpsi     {{\ensuremath{{\PJ\mskip -3mu/\mskip -2mu\Ppsi}}}\xspace}

\def\Y#1S{\ensuremath{\PUpsilon{(#1S)}}\xspace}

\def\LorLbar     {\kern \thebaroffset\optbar{\kern -\thebaroffset \PLambda}\xspace}

\newcommand{\decay}[2]{\ensuremath{#1\!\to #2}\xspace} 

\def\to                 {\ensuremath{\rightarrow}\xspace}

\def\AT#1     {\ensuremath{A_{\mathrm{T}}^{#1}}\xspace}           

\def\C#1      {\ensuremath{\mathcal{C}_{#1}}\xspace}                       
\def\Cp#1     {\ensuremath{\mathcal{C}_{#1}^{'}}\xspace}                    
\def\Ceff#1   {\ensuremath{\mathcal{C}_{#1}^{\mathrm{(eff)}}}\xspace}        
\def\Cpeff#1  {\ensuremath{\mathcal{C}_{#1}^{'\mathrm{(eff)}}}\xspace}       
\def\Ope#1    {\ensuremath{\mathcal{O}_{#1}}\xspace}                       
\def\Opep#1   {\ensuremath{\mathcal{O}_{#1}^{'}}\xspace}                    

\newcommand{\nospaceunit}[1]{\ensuremath{\text{#1}}}       
\newcommand{\aunit}[1]{\ensuremath{\text{\,#1}}}       

\newcommand{\tev}{\aunit{Te\kern -0.1em V}\xspace}
\newcommand{\gev}{\aunit{Ge\kern -0.1em V}\xspace}
\newcommand{\mev}{\aunit{Me\kern -0.1em V}\xspace}
\newcommand{\kev}{\aunit{ke\kern -0.1em V}\xspace}
\newcommand{\ev}{\aunit{e\kern -0.1em V}\xspace}
 
\newcommand{\mevc}{\ensuremath{\aunit{Me\kern -0.1em V\!/}c}\xspace}
\newcommand{\gevc}{\ensuremath{\aunit{Ge\kern -0.1em V\!/}c}\xspace}
\newcommand{\mevcc}{\ensuremath{\aunit{Me\kern -0.1em V\!/}c^2}\xspace}
\newcommand{\gevcc}{\ensuremath{\aunit{Ge\kern -0.1em V\!/}c^2}\xspace}

\def\mum  {\ensuremath{\,\upmu\nospaceunit{m}}\xspace}

\def\fb   {\ensuremath{\aunit{fb}}\xspace}
\def\invfb   {\ensuremath{\fb^{-1}}\xspace}

\def\gsim{{~\raise.15em\hbox{$>$}\kern-.85em
          \lower.35em\hbox{$\sim$}~}\xspace}
\def\lsim{{~\raise.15em\hbox{$<$}\kern-.85em
          \lower.35em\hbox{$\sim$}~}\xspace}

\def\sPlot{\mbox{\em sPlot}\xspace}

\def\pt         {\ensuremath{p_{\mathrm{T}}}\xspace}

\def\ptot       {\ensuremath{p}\xspace}

\def\evtgen     {\mbox{\textsc{EvtGen}}\xspace}

\def\geant      {\mbox{\textsc{Geant4}}\xspace}

\def\photos     {\mbox{\textsc{Photos}}\xspace}

\def\pythia     {\mbox{\textsc{Pythia}}\xspace}

\def\tell1  {TELL1\xspace}
\def\ukl1   {UKL1\xspace}

\newcommand{\lhcborcid}[1]{\href{https://orcid.org/#1}{\hspace*{0.1em}\raisebox{-0.45ex}{\includegraphics[width=1em]{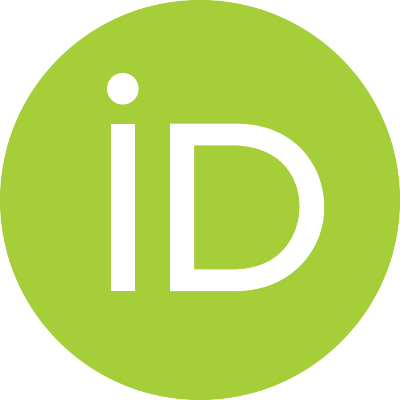}}}}

\def\jpsitomm     {\decay{\jpsi}{\mup\mun}}
\def\jpsitofourm     {\decay{\jpsi}{\mup\mun\mup\mun}}

\usepackage{cite} 
\usepackage{mciteplus}

\usepackage{longtable} 

\begin{document}

\renewcommand{\thefootnote}{\fnsymbol{footnote}}
\setcounter{footnote}{1}


\begin{titlepage}
\pagenumbering{roman}

\vspace*{-1.5cm}
\centerline{\large EUROPEAN ORGANIZATION FOR NUCLEAR RESEARCH (CERN)}
\vspace*{1.5cm}
\noindent
\begin{tabular*}{\linewidth}{lc@{\extracolsep{\fill}}r@{\extracolsep{0pt}}}
\ifthenelse{\boolean{pdflatex}}
{\vspace*{-1.5cm}\mbox{\!\!\!\includegraphics[width=.14\textwidth]{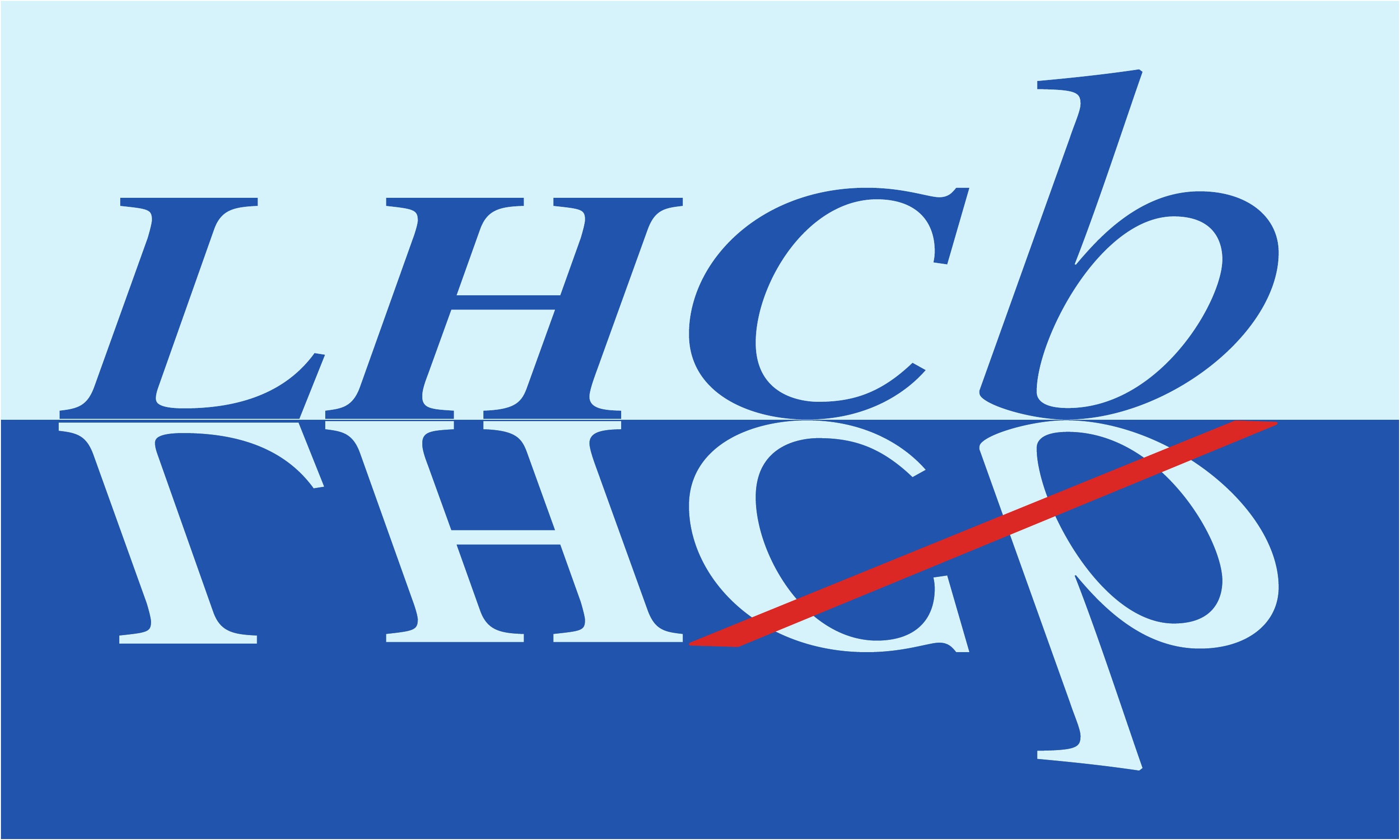}} & &}%
{\vspace*{-1.2cm}\mbox{\!\!\!\includegraphics[width=.12\textwidth]{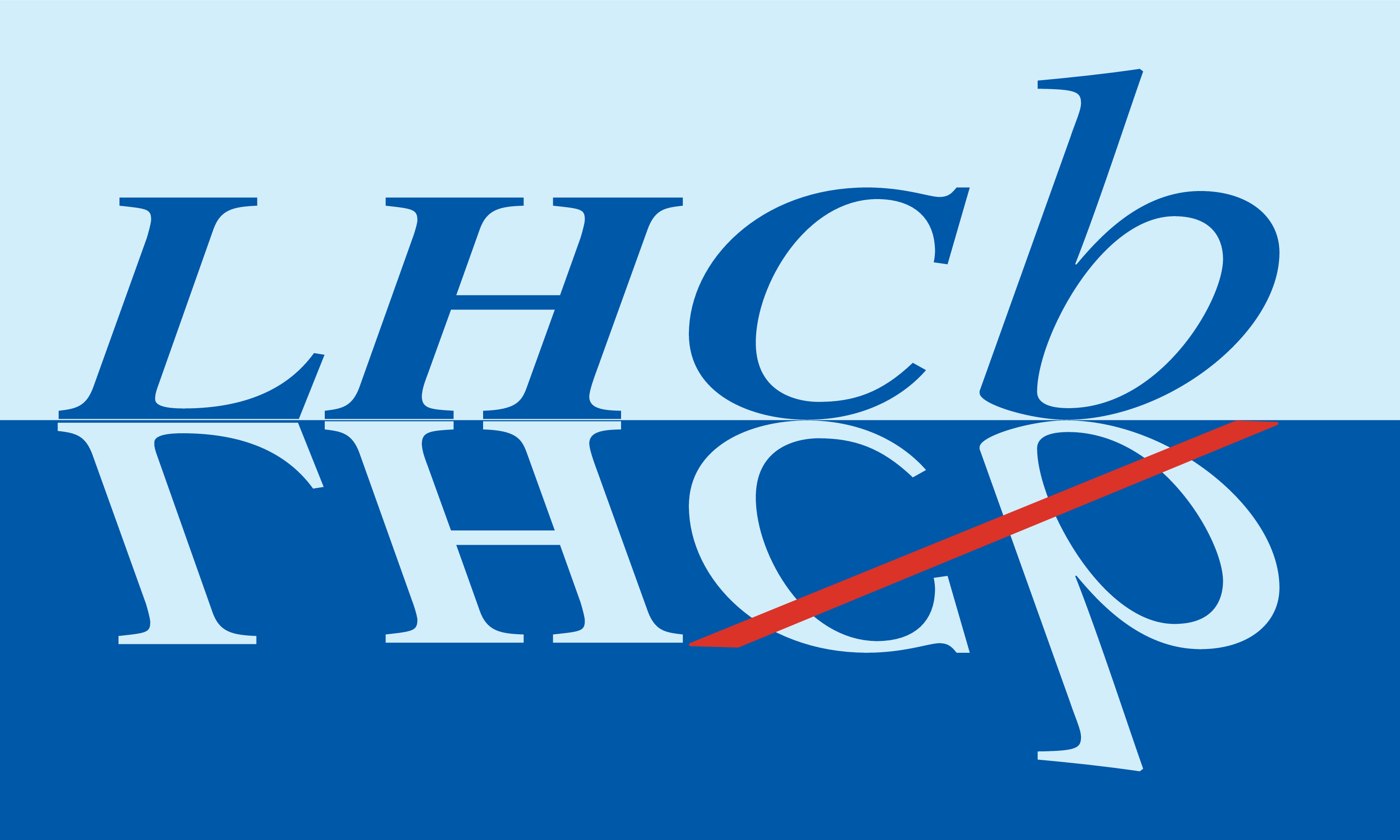}} & &}%
\\
 & & CERN-EP-2024-201 \\  
 & & LHCb-PAPER-2024-016 \\  
 & & 22 January 2025 \\
 & & \\
\end{tabular*}

\vspace*{4.0cm}

{\normalfont\bfseries\boldmath\huge
\begin{center}
  \papertitle 
\end{center}
}

\vspace*{2.0cm}

\begin{center}
\paperauthors\footnote{Authors are listed at the end of this paper.}
\end{center}

\vspace{\fill}

\begin{abstract}
  \noindent
The rare electromagnetic $J\mskip -3mu/\mskip -2mu\psi \to \mu^+\mu^-\mu^+\mu^-$ decay is observed with a significance greatly exceeding the discovery threshold, using proton-proton collision data collected by the LHCb experiment during 2016--2018 at a center-of-mass energy of 13\tev, corresponding to an integrated luminosity of  $5.4\,\text{fb}^{-1}$. The rate of this decay is measured relative to that of the $J\mskip -3mu/\mskip -2mu\psi \to \mu^+\mu^-$ mode. Using the QED model for the four-muon decay in the efficiency estimation, its branching fraction is determined to be 
\begin{equation*}
   {{\cal{B}}}(J\mskip -3mu/\mskip -2mu\psi \to \mu^+\mu^-\mu^+\mu^-) = (1.13\pm0.10\pm0.05\pm0.01)\times 10^{-6},
\end{equation*}
where the uncertainties are statistical, systematic and due to the uncertainty on the branching fraction of the $J\mskip -3mu/\mskip -2mu\psi \to \mu^+\mu^-$ decay.
\end{abstract}

\vspace*{2.0cm}

\begin{center}
Published in JHEP 12 (2024) 062
\end{center}

\vspace{\fill}

{\footnotesize 
\centerline{\copyright~\papercopyright. \href{\paperlicenceurl}{\paperlicence}.}}
\vspace*{2mm}

\end{titlepage}


\newpage
\setcounter{page}{2}
\mbox{~}
%
%
%
%


\renewcommand{\thefootnote}{\arabic{footnote}}
\setcounter{footnote}{0}

\cleardoublepage


\pagestyle{plain} 
\setcounter{page}{1}
\pagenumbering{arabic}


\section{Introduction}
\label{sec:Introduction}

Decays of heavy-flavour mesons into final states containing more than two leptons are an interesting but rather unexplored probe of the Standard Model (SM). The SM rates of such decays can be computed relatively precisely. In addition, possible physics beyond the SM that couples to leptons, may contribute to the rate of these decay processes~\cite{Chala:2019vzu}. 

Four-lepton decay modes have been observed for the light pseudoscalar \piz and \etaz mesons~\cite{PDG2024}, where they proceed through a diphoton diagram. 
Recently, the \besiii collaboration reported observation of the $\etapr\to\epem\epem$ decay~\cite{BESIII:2022cul}, and the \cms collaboration has observed the $\etaz\to\mumu\mumu$ decay~\cite{CMS:2023gui}. In addition, the LHCb collaboration has searched for the loop-level four-muon decay of neutral kaons~\cite{LHCb-PAPER-2022-035}. 
Besides these low-mass particles, four-lepton final states are being actively explored for much heavier particles, with electroweak and Higgs boson measurements~\cite{PDG2024}. However, knowledge of four-lepton decays of heavy-quark hadrons is limited, and sufficiently large datasets required for their study only became available recently. The LHCb experiment has searched for the four-muon decays of \Bds mesons~\cite{LHCb-PAPER-2021-039}, which proceed through  loop-level electroweak diagrams involving the initial- or final-state radiation of a dimuon, but no signal was observed. However, studies of tree-level multilepton decays of heavy-quark hadrons, such as $\jpsitofourm$, have not been performed until recently. 
As these decays are comparatively abundant, their study would also allow for validation of the theoretical description of final-state radiation (FSR) which can then be applied to the much rarer four-lepton decays of $B$~mesons~\cite{Danilina:2018uzr,LHCb-PAPER-2021-039}. 

Recently, theoretical calculations for the electromagnetic four-lepton decay rates of the vector quarkonia states, which are within the reach of current experiments, have been published~\cite{Chen:2020bju}. Of special interest is the \jpsi meson, a charmonium state with the quantum numbers $J^{PC} = 1^{--}$, that has a large production cross-section in various experimental environments. It has a large electromagnetic decay width compared to its strong decay width, which is suppressed due to the Okubo–Zweig–Iizuka rule~\cite{Kopke:1988cs,Voloshin:2007dx}. This makes the four-lepton decay of the \jpsi meson an excellent tool for probing rare electromagnetic processes and the theory of quantum electrodynamics (QED) at the LHC. 

The predicted branching fractions, $\mathcal{B}$, in Ref.~\cite{Chen:2020bju} are \mbox{$(52.88\pm0.28)\times 10^{-6}$}, \mbox{$(36.73\pm0.20)\times 10^{-6}$} and \mbox{$(0.974\pm0.005)\times 10^{-6}$} for \mbox{$\jpsi \to \epem\epem$}, \mbox{$\jpsi \to \epem\mumu$} and \mbox{$\jpsi \to \mumu\mumu$}, respectively, where the uncertainty comes only from that on the known value of $\mathcal{B}(\jpsitomm)$. The difference in rates between decays to leptons of different flavours is due to the mass difference between electrons and muons. 

The \besiii experiment has searched for the four-lepton decays of the \jpsi meson~\cite{BESIII:2021ocn}, observing its decays to the $\epem\epem$ and $\epem\mumu$ final states, with measured branching fractions being consistent with the theoretical predictions presented above. 
The four-muon decay was not observed owing to its smaller rate, and only an upper limit was set on its branching fraction, $\mathcal{B}(\jpsitofourm)<1.6\times 10^{-6}$ at 90\% confidence level. 
Recently, the \cms experiment has observed the $\jpsi\to\mumu\mumu$ decay, using proton-proton ($pp$) collision data corresponding to an integrated luminosity of 33.6\invfb and reports a signal yield of $11.6^{+3.8}_{-3.1}$. The corresponding branching fraction was measured to be $[1.01^{+0.33}_{-0.27}\,\text{(stat)} \pm 0.04\,\text{(syst)}]\times 10^{-6}$, assuming a phase-space decay model for the efficiency estimation~\cite{CMS:2024jpsi}. 

Large production cross-sections of vector quarkonia states at the \lhcb experiment~\cite{LHCb-PAPER-2015-037,LHCb-PAPER-2018-002} make it a natural place for precision measurements of four-lepton quarkonia decays.  
Since most of the \ccbar and all \bbbar mesons are produced promptly in the $pp$ collision, reconstructed samples of heavy quarkonia suffer from very large backgrounds, making prompt decays hard to isolate. 
An alternative mechanism of \ccbar production is via decays of \bquark hadrons, referred to as secondary production, in which the production vertex of \ccbar mesons is displaced from the $pp$ interaction point. 
At LHCb, electrons are more difficult to reconstruct and identify than muons, resulting in both poorer efficiency and worse mass resolution in decay modes with electrons. Consequently, the $\jpsi \to \mu^+\mu^-\mu^+\mu^-$ decay mode  is the ideal candidate for a first study of four-lepton decays of vector quarkonia at \lhcb.

\begin{figure}[!tb]
        \centering

        \includegraphics[height=1.8in]{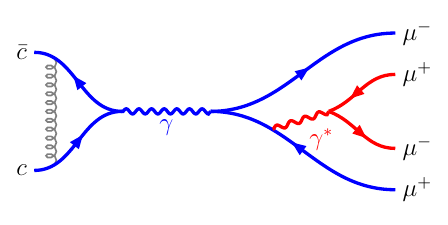}

\caption{Tree-level diagram for the four-muon decay of $J\mskip -3mu/\mskip -2mu\psi$ mesons. The virtual photon $\gamma^*$ can be emitted from either muon leg. The blue part of the diagram corresponds to the $J\mskip -3mu/\mskip -2mu\psi \to \mu^+\mu^-$ decay.}
\label{diag:jpsi24l}
\end{figure}

The $\jpsitofourm$ decay proceeds predominantly via final-state radiation of a virtual photon $\gamma^*$, as shown in Fig.~\ref{diag:jpsi24l}. The analogous initial-state radiation process is not expected to contribute significantly, as the process $\jpsi \to \gamma\gamma$ is forbidden by the requirement of $C$-parity conservation in electromagnetic processes. 
The rate of the FSR process is suppressed by two factors of the electromagnetic coupling constant compared to the dimuon decay without FSR. 
Further suppression arises from the relatively large muon mass. 
Finally, the presence of identical muons in the final state leads to interference terms between possible diagrams.  
It should be noted that most of the processes with $\jpsi \to VV \to \mumu\mumu$ where $V$ is a light-quark vector resonance, that could lead to the same final state, are forbidden by $C$-parity conservation in the strong and electromagnetic interactions. 

Assuming that the SM rate is dominated by the FSR process, it is possible to separate the hadronic physics of the \jpsi annihilation into a virtual photon (identical for the \jpsitofourm and \jpsitomm decays), and the FSR process (only appearing in the \jpsitofourm decay). 
Therefore, the ratio of branching fractions of \jpsitofourm and \jpsitomm decays is an observable that can be theoretically predicted to a rather high precision. This observable is also convenient experimentally, allowing the cancellation of several sources of systematic uncertainty. The ratio of branching fractions is defined as
\begin{equation}
    R_{\mathcal{B}} \equiv \frac{\mathcal{B}(\jpsitofourm)}{\mathcal{B}(\jpsitomm)}.
\end{equation}
Throughout this paper, \jpsitofourm will be referred to as the ``signal'' mode and \jpsitomm as the ``normalisation'' mode.

The remainder of this document is organised as follows. Section~\ref{sec:DataSim} contains a short description of the LHCb detector, the data processing chain and the simulation. The candidate selection and background suppression is described in Sec.~\ref{sec:Selection}. The efficiency modelling is described in Sec.~\ref{sec:effs}, and the invariant-mass fits used to extract the final results in Sec.~\ref{sec:fits}. Systematic uncertainties are discussed in Sec.~\ref{sec:syst}. This paper concludes with a discussion of the results in Sec.~\ref{sec:Conclusion}.

\section{LHCb detector, datasets and simulation}
\label{sec:DataSim}

The \lhcb detector~\cite{LHCb-DP-2008-001,LHCb-DP-2014-002} is a single-arm forward
spectrometer covering the \mbox{pseudorapidity} range $2<\eta <5$,
designed for the study of particles containing \bquark or \cquark
quarks. The detector includes a high-precision tracking system
consisting of a silicon-strip vertex detector surrounding the $pp$
interaction region, a large-area silicon-strip detector located
upstream of a dipole magnet with a bending power of about
$4{\mathrm{\,T\,m}}$, and three stations of silicon-strip detectors and straw
drift tubes placed downstream of the magnet.
The tracking system provides a measurement of the momentum, \ptot, of charged particles with a relative uncertainty that varies from 0.5\% at low momentum to 1.0\% at 200\gevc.
The minimum distance of a track to a primary $pp$ collision vertex (PV), the impact parameter, is measured with a resolution of $(15+29/\pt)\mum$,
where \pt is the component of the momentum transverse to the beam, in\,\gevc.
The particle identification is provided by several subdetectors. Two ring-imaging Cherenkov detectors mainly aim at distinguishing different types of charged hadrons, but also contribute to muon and electron identification at low momenta. A calorimeter system consisting of scintillating-pad and preshower detectors, an electromagnetic and a hadronic calorimeter, allows one to distinguish photons, electrons, hadrons and muons. Muons are identified by a system composed of alternating layers of iron and multiwire proportional chambers. This system returns a binary value for positive muon identification if the muon candidate produces signal in a predefined number of stations that depends on its momentum~\cite{LHCb-DP-2013-001}.
Furthermore, a multivariate classifier combines information from each detector system, returning a single probability value, denoted here as $P_\mu$, for the muon hypothesis~\cite{LHCb-DP-2018-001}. 
This measurement relies on data collected by the \lhcb detector during 2016--2018, in $pp$ collisions at a center-of-mass energy of $13\tev$, corresponding to an integrated luminosity of 5.4\,$\text{fb}^{-1}$. 
The online event selection is performed by a trigger~\cite{LHCb-DP-2012-004}, 
which consists of a hardware stage, based on information from the muon systems and the calorimeters, followed by a software stage, which applies a full event
reconstruction.

Simulation of the decays of interest is required to model the effects of the detector acceptance and the imposed selection requirements. 
In the simulation, $pp$ collisions are generated using \pythia~\cite{Sjostrand:2007gs,*Sjostrand:2006za} with a specific \lhcb configuration~\cite{LHCb-PROC-2010-056}.
Decays of unstable particles are described by \evtgen~\cite{Lange:2001uf}, where on-shell final-state radiation is generated using \photos~\cite{davidson2015photos}.
The interaction of the generated particles with the detector, and its response,  are implemented using the \geant toolkit~\cite{Allison:2006ve, *Agostinelli:2002hh} as described in Ref.~\cite{LHCb-PROC-2011-006}. 

The \jpsitomm decay is simulated using the \evtgen model for decays of unpolarised vector mesons into two leptons. 
Simulated \jpsitofourm decays are generated according to an \evtgen model based on the analytical leading-order calculation of the QED decay rate. 
This model is adapted from that used by the \besiii experiment in Ref.~\cite{BESIII:2021ocn} by assuming an unpolarised \jpsi production.

The simulation of \jpsitomm and \jpsitofourm decays undergoes several offline calibration steps to improve its agreement with data. Muon identification is calibrated by resampling the distributions of $P_\mu$ for each muon, using kernel density estimation. 
This approach relies on large tag-and-probe calibration datasets of \jpsitomm decays, collected in parallel to the regular data taking~\cite{LHCb-DP-2018-001}. 
Dedicated \jpsitomm datasets are used to calibrate the muon tracking efficiency and apply a correction weight per track to simulated events~\cite{LHCb-DP-2013-002}. The hardware trigger response is also calibrated using the tag-and-probe technique relying on \jpsitomm datasets where the trigger decision was taken using particles other than muons from the \jpsi decay~\cite{LHCb-PUB-2014-039}. 
As mentioned above, the two dominant production mechanisms of \jpsi mesons are prompt production in the $pp$ collision, or secondary production in decays of \bquark hadrons.
The proportion of \jpsi candidates produced through the prompt and secondary mechanisms in simulation is weighted to match the value measured at LHCb~\cite{LHCb-PAPER-2015-037}. 
Finally, the kinematics of the \jpsi candidates is weighted in simulation in bins of $\pt(\jpsi)$ and $\eta(\jpsi)$, separately for prompt and secondary production mechanisms, to match those of the \jpsitomm data; these weights are also applied to the four-muon datasets. 

\section{Candidate selection}
\label{sec:Selection}

The data are analysed separately in two disjoint categories, prompt and \mbox{secondary}. 
The prompt category contains \jpsi candidates whose decay-length significance with respect to the PV (defined as the distance between the PV and the \jpsi decay vertex divided by its uncertainty) does not exceed three standard deviations, while \jpsi candidates with decay-length significance above this threshold are assigned to the secondary category. The prompt category is characterised by a large background from particles other than muons, consisting mostly of hadrons, produced directly in the $pp$ collisions. They can be misidentified as muons due to their decay in flight or due to assigning random hits in the muon system to their track. In addition, some energetic hadrons are also able to pass through the calorimeters and reach the muon system. 
The secondary category offers less challenging background conditions, therefore an equivalent signal purity can be achieved with a looser selection.  
It also benefits from the trigger system of the experiment, which is optimised for weak decays of \bquark and \cquark hadrons. 

At the hardware trigger stage, for both the signal and normalisation decays, one of the muons is required to have \pt greater than 1.3--1.8\gevc, depending on the data-taking period. 
At the first software trigger stage, at least one dimuon candidate is required in the datasets of the prompt category, while at least one dimuon or a displaced muon is required in the datasets of the secondary category. 
At the second software trigger stage, a four-muon candidate is required in the signal dataset of the prompt category, while a displaced two-, three- or four-muon secondary vertex with a significant displacement from any PV is required in the signal dataset of the secondary category, as explained in Ref.~\cite{LHCb-DP-2019-001}. For the normalisation dataset, a dimuon candidate with invariant mass consistent with the known \jpsi mass~\cite{PDG2024} is required at the second software trigger stage.  
All track candidates used in this analysis are required to be positively identified as muons. 
This implies that their momentum has to exceed 3\gevc, which is the threshold to pass through the calorimeters and reach the muon system. 

Due to the high multiplicity of prompt tracks in an average $pp$ collision, it is computationally challenging to fit all possible four-track vertices directly.
The four-muon dataset of the prompt category is therefore constructed by forming a good vertex from two opposite-charge dimuon candidates, each with a good vertex quality. Both the dimuon and four-muon candidates are required to have a decay-length significance less than three standard deviations, in order to prevent overlap with the secondary category. 
Each muon candidate is required to have  \pt greater than 500\mevc and $p$ greater than 10\gevc. 

The four-muon candidates of the secondary category are reconstructed in a different manner, by directly forming a good-quality four-track vertex, displaced from the associated PV by more than three standard deviations. 
At least one muon candidate must have \pt greater than 500\mevc, and at least one other must have \pt greater than 300\mevc. 

The \jpsitomm candidates of the prompt category are required to have two oppositely charged muons with \pt above 650\mevc and $p$ above 10\gevc that form a good vertex. The transverse momentum of the dimuon must exceed 3\gevc. The \jpsitomm candidates of the secondary category are constructed from two oppositely charged muons with \pt above 500\mevc, forming a good-quality vertex that is displaced from the PV.

To reject fake tracks sharing track segments with genuine muons, the angle between any two muon tracks in the laboratory frame is required to exceed 0.5\,mrad. In the four-muon dataset, all opposite-charge dimuon combinations are required to have an invariant mass below 2900\mevcc in order to suppress contributions from \jpsitomm decays combined with two other muons. For the \jpsitomm decay, no further selection is required. Fewer than one per mille dimuon events have more than one \jpsitomm candidate, and all candidates are retained, as pair production of \jpsi mesons is not infrequent~\cite{LHCb-PAPER-2023-022}. 

In the four-muon dataset, the dominant background source is the random combination of muons and misidentified hadrons. A multivariate classifier is used to suppress this background. 
It is based on the boosted decision tree~(BDT)  algorithm with gradient boost~\cite{Breiman,AdaBoost}, as implemented in the TMVA
toolkit~\cite{Hocker:2007ht,*TMVA4}, and is trained separately for the prompt and secondary categories of \jpsitofourm decays. The training uses the following information: \pt, $\eta$ and the vertex quality of the four-muon candidate, as well as the minimum and maximum $P_\mu$ values for the muon hypothesis amongst the four muon candidates. In the secondary category, in addition, the angle between the momentum vector of the \jpsi candidate and the vector joining the PV and the \jpsi decay vertex, is used.  
As a signal proxy, simulated \jpsitofourm decays are used, with all calibrations applied, as discussed above. As a background proxy, four-muon candidates in data are selected in the invariant-mass sidebands, \mbox{$2700 <m(\mumu\mumu)< 2900 \mevcc$} and \mbox{$3200<m(\mumu\mumu)<3400 \mevcc$}.
The requirement on the BDT output value is optimised by maximising the quantity $S/\sqrt{S+B}$ separately for the prompt and secondary categories. Here, $S$ is the expected \mbox{\jpsitofourm} yield passing a certain BDT requirement, estimated assuming the predicted branching fraction value from Ref.~\cite{Chen:2020bju}, and $B$ is the background yield in the signal region. The latter is estimated from the observed background yield in the sidebands of the signal peak. The resulting BDT selection requirements retain about 50\% (60\%) of signal \jpsitofourm candidates in the prompt (secondary) dataset, while rejecting 96\% (98\%) of background candidates. 

A possible contamination due to misidentified hadrons in four-body decays of quarkonia, such as the $\jpsi \to \pip\pim\pip\pim$ decay or the as yet unobserved $\jpsi \to \pip\pim\mumu$ decay, is tested by requiring a positive pion or kaon identification on at least two muons, as well as by loosening the muon identification criteria. No hint of such backgrounds is seen. 

As the \jpsitofourm decay is rare, it is unlikely to occur twice in the same event even in case of \jpsi pair production. 
However, given that the prompt four-muon candidates are built from two dimuon candidates, constructed independently, each four-muon combination is reconstructed twice, unless one of these two candidates fails the dimuon vertex-quality requirement. These two candidates are constructed from the same four final-state tracks but may differ  slightly due to the vertex reconstruction. 
Therefore, only one candidate per event is arbitrarily selected to be retained, both in data and simulation, after the full selection is applied. 
In the four-muon candidates of the secondary category, no duplication occurs at the reconstruction stage, and less than 1\% of events contain multiple candidates. Where multiple candidates exist, one is selected arbitrarily. 

\section{Efficiency estimation}
\label{sec:effs}

In order to measure the value of $R_{\mathcal{B}}$, the ratio of the observed \jpsitofourm and \jpsitomm yields needs to be corrected for the total detection efficiency for each of the two decay modes. 
The efficiency estimation is performed using a calibrated simulation of \jpsi decays with the prompt and secondary categories merged in the denominator of the efficiency definition. The efficiency of the decay-length significance requirements that define the two analysis categories is therefore part of the total efficiency for each category. 
This aims for a consistent treatment of simulation and data: although in simulation it is possible to fully distinguish the \jpsi candidates produced promptly and in $b$-hadron decays using truth-level information, this cannot be done in data and is not equivalent to the analysis category definition based on the decay-length significance which is a reconstructed quantity. 

\begin{table}[!b]
  \centering
  \caption{Total efficiency for the \jpsitofourm and \jpsitomm decays, as well as their ratio, split by  analysis category. The uncertainties only account for the size of the simulated samples.} 
  \label{tab:total_effs}
  \begin{tabular}{c c c c}
Category & $\varepsilon(\jpsitofourm)$  & $\varepsilon(\jpsitomm)$ & Ratio   \\
\hline
Prompt & $(0.052 \pm 0.001)\%$ & $(1.992\pm0.001)\%$ & $0.0262\pm0.0002$ \\
Secondary & $(0.085\pm0.001)\%$ & $(0.518\pm0.001)\%$ & $0.164\pm0.001$\\
 \end{tabular}
\end{table}

\begin{figure}[!tb]
        \centering

        \includegraphics[height=5.1in]{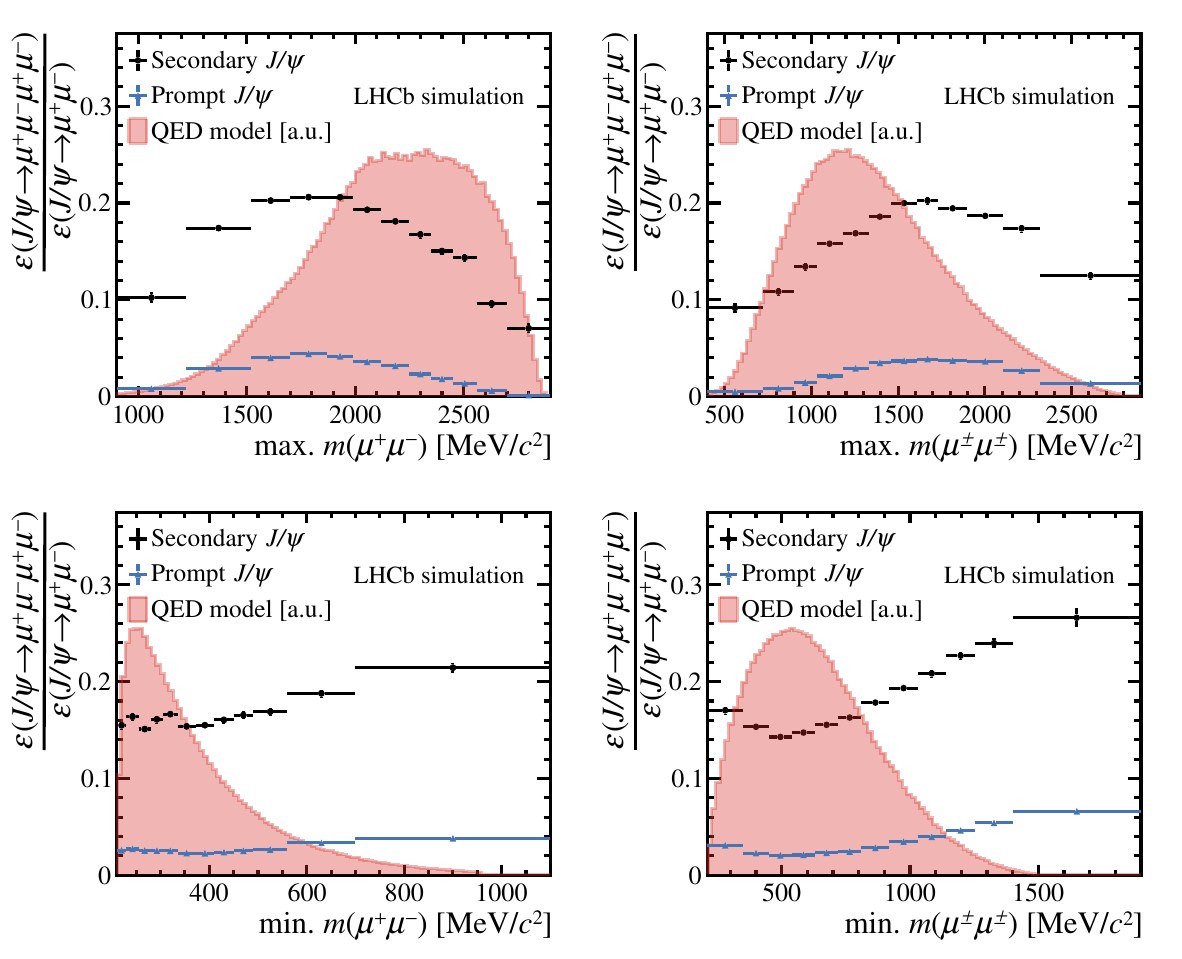}

    \caption{Efficiency of the $J\mskip -3mu/\mskip -2mu\psi \to \mu^+\mu^-\mu^+\mu^-$ selection as a function of the maximum and minimum values of the invariant masses of the same-charge or opposite-charge muon pairs, for the prompt and secondary categories of $J\mskip -3mu/\mskip -2mu\psi$ candidates, normalised to the total efficiency of the corresponding $J\mskip -3mu/\mskip -2mu\psi \to \mu^+\mu^-$ category. The underlying distribution of these variables in the QED model prior to detector effects is shown for reference, with an arbitrary normalisation.}
    \label{fig:eff_trends}
\end{figure}

The total efficiency accounts for the geometric and kinematic acceptance of the LHCb detector, and trigger, reconstruction and selection requirements.  
The values of total efficiency, as well as the efficiency ratio between the \jpsitofourm and \jpsitomm datasets, are shown in Table~\ref{tab:total_effs}, where the quoted uncertainty is only due to the size of simulated samples, discussed later. 
The efficiency ratio is lower in the prompt category, due to the stringent requirements imposed in the trigger and offline selection for the prompt four-muon dataset. 
It is worth pointing out that the weighting of the \jpsi kinematics changes the efficiency ratio by about 10\% (30\%) in the secondary (prompt) categories. This is because simulation underestimates the fraction of low-momentum \jpsi mesons, for which the efficiency is typically low, especially in the prompt category. 
A large shift in efficiencies, exceeding 30\%, is also observed when comparing the simulation based on the QED calculation with less accurate models, such as those with kinematics uniformly distributed across the phase space. This is due to the fact that the decay model choice impacts the simulated muon kinematics, and therefore, the efficiency of the selection requirements. 
Indeed, the ratio of efficiencies shown in  Fig.~\ref{fig:eff_trends} exhibits dependence on the dimuon invariant-mass observables in the \jpsitofourm decay. 
Therefore, a simultaneous analysis of the prompt and secondary categories is a powerful tool to cross-check  the obtained results. 

\section{Mass modelling}
\label{sec:fits}

Signals are separated from the residual background with fits to the four-muon (dimuon) invariant mass for the signal (normalisation) datasets. 
An extended maximum-likelihood fit is performed separately for each studied dataset. It is implemented in the \textsc{RooFit} toolkit~\cite{Verkerke:2003ir} within the \textsc{ROOT} framework~\cite{Brun:1997pa}.

\begin{figure}[!tb]
        \centering
    \begin{subfigure}{0.5\textwidth}
        \centering
        \includegraphics[height=2.8in]{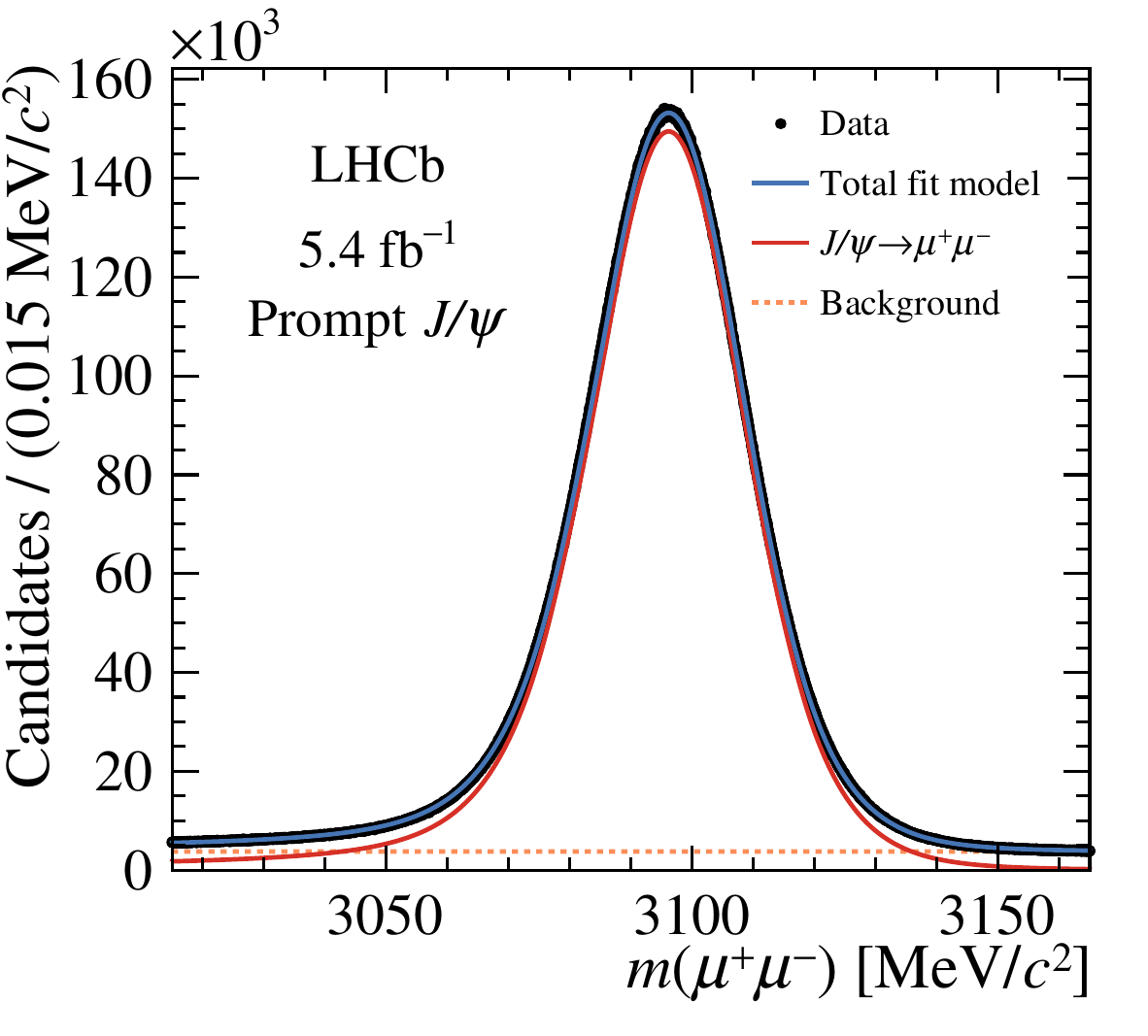}
    \end{subfigure}%
    \begin{subfigure}{0.5\textwidth}
        \centering
        \includegraphics[height=2.8in]{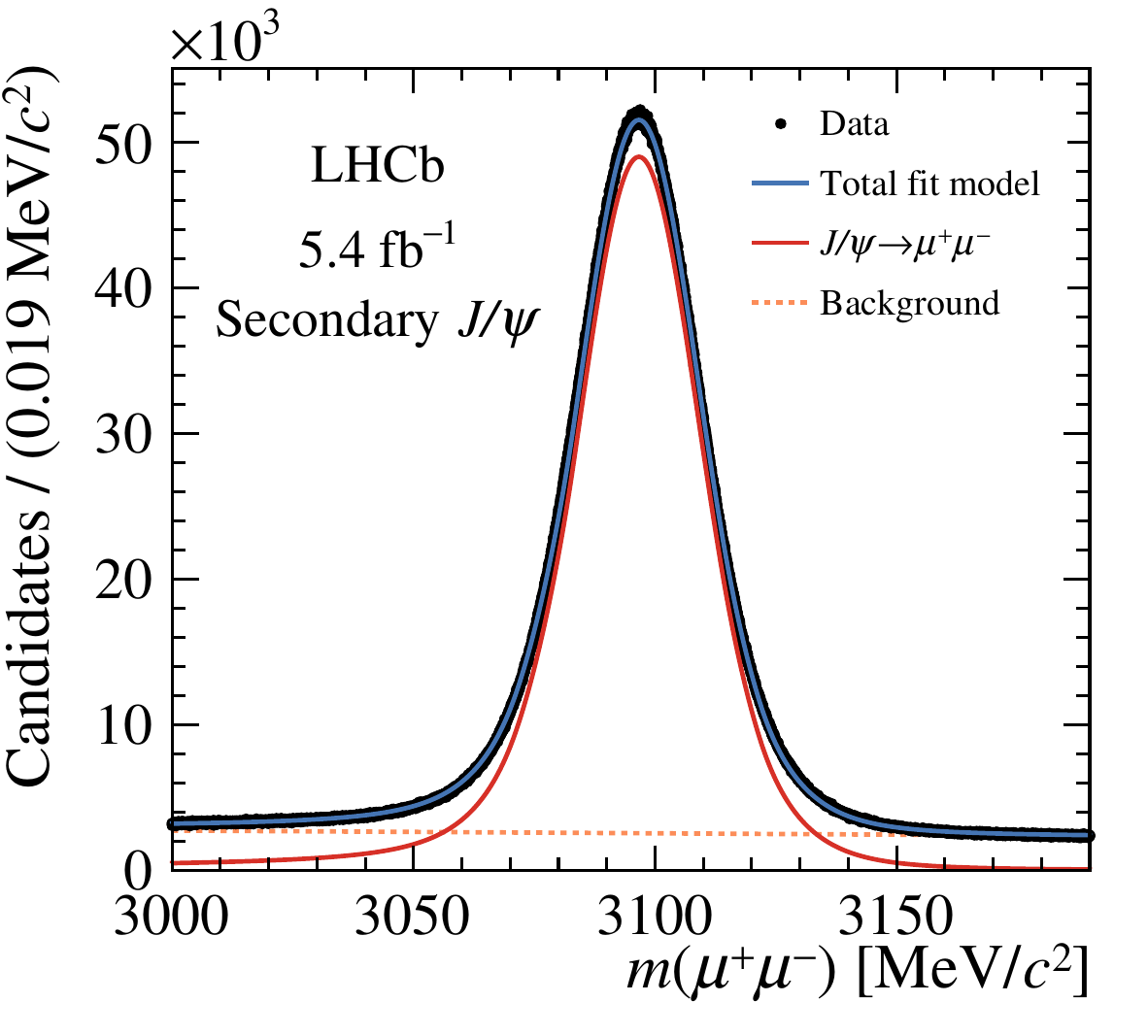}
    \end{subfigure}

    \caption{Invariant-mass distributions for (left) prompt $J\mskip -3mu/\mskip -2mu\psi \to \mu^+\mu^-$ candidates and (right) secondary $J\mskip -3mu/\mskip -2mu\psi \to \mu^+\mu^-$ candidates. The results of the fit are overlaid.}
    \label{fig:Jpsi22mfit}
\end{figure}

First, fits to the invariant-mass distributions of the selected dimuon candidates are performed. 
The signal shape is parameterised with a linear combination of two Crystal Ball functions~\cite{Skwarnicki:1986xj} modified to have power-law tails on both sides of the peak, and one Gaussian function. The two Crystal Ball functions are required to share the same mean.
The shape parameters are determined by fitting the simulated datasets. In the fit to data, the relative fractions of individual components of the signal shape, and all the power-law parameters of the Crystal Ball functions, are fixed to the values obtained from simulation. The mean of one Crystal Ball function and its resolution parameter are allowed to vary, as simulation slightly underestimates the resolution. The difference between the means of the Gaussian and Crystal Ball functions is fixed from simulation, as are the ratios of the width parameters of these three components. The combinatorial background is modelled by a linear function with a slope that is allowed to vary.
Due to the large size of the dimuon datasets, binned maximum-likelihood fits are performed with 10\,000 bins. 
The resulting invariant-mass distributions in the prompt and secondary \jpsitomm categories, as well as the fit results, are shown in Fig.~\ref{fig:Jpsi22mfit}: 
$\left(341\,880 \pm 21\right)\times 10^{3}$ \jpsitomm decays are observed in the prompt category, and $\left(91\,704\pm 11\right)\times 10^{3}$ in the secondary category, where uncertainties are statistical only.

\begin{figure}[!tb]
        \centering
    \begin{subfigure}{0.5\textwidth}
        \centering
        \includegraphics[height=2.8in]{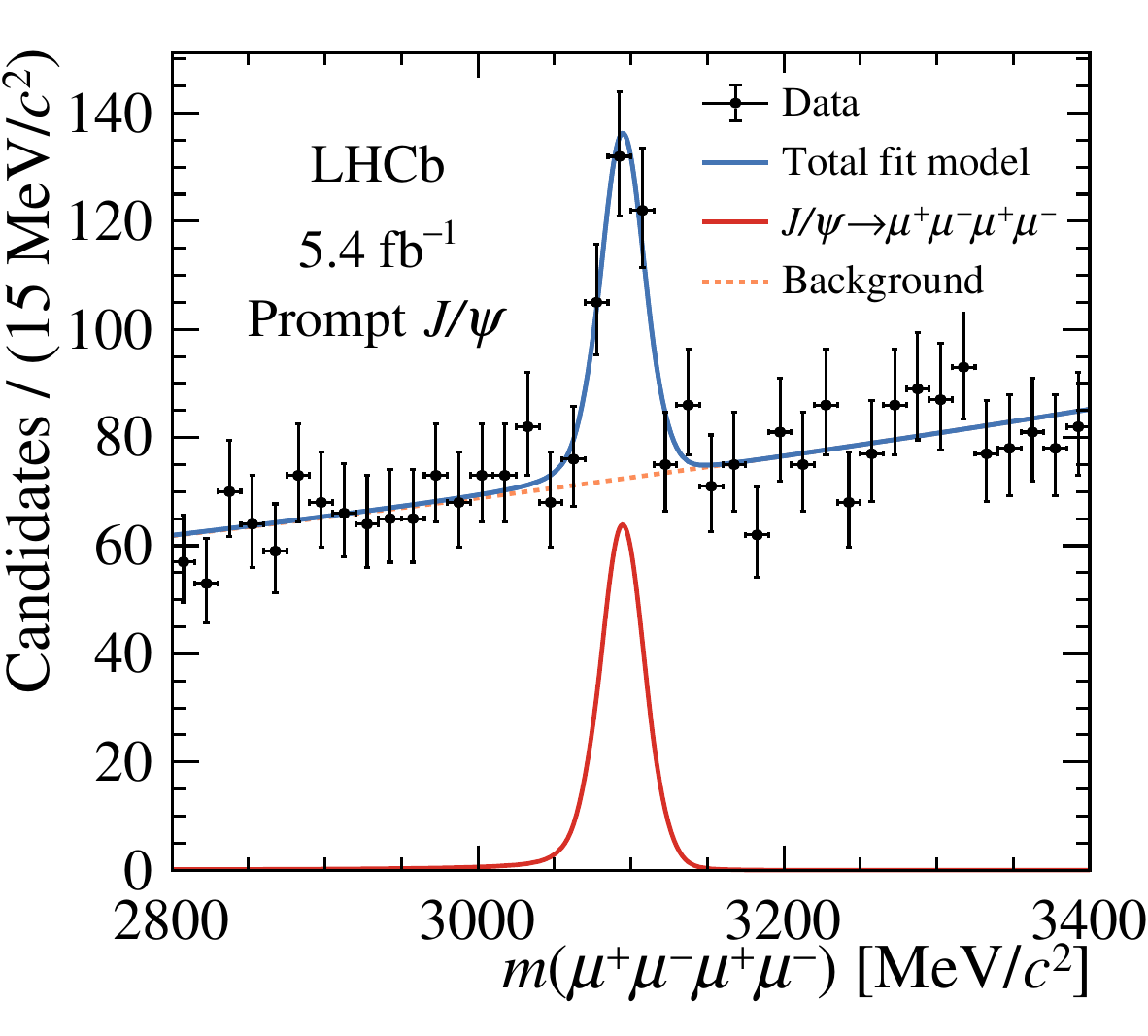}
    \end{subfigure}%
    \begin{subfigure}{0.5\textwidth}
        \centering
        \includegraphics[height=2.8in]{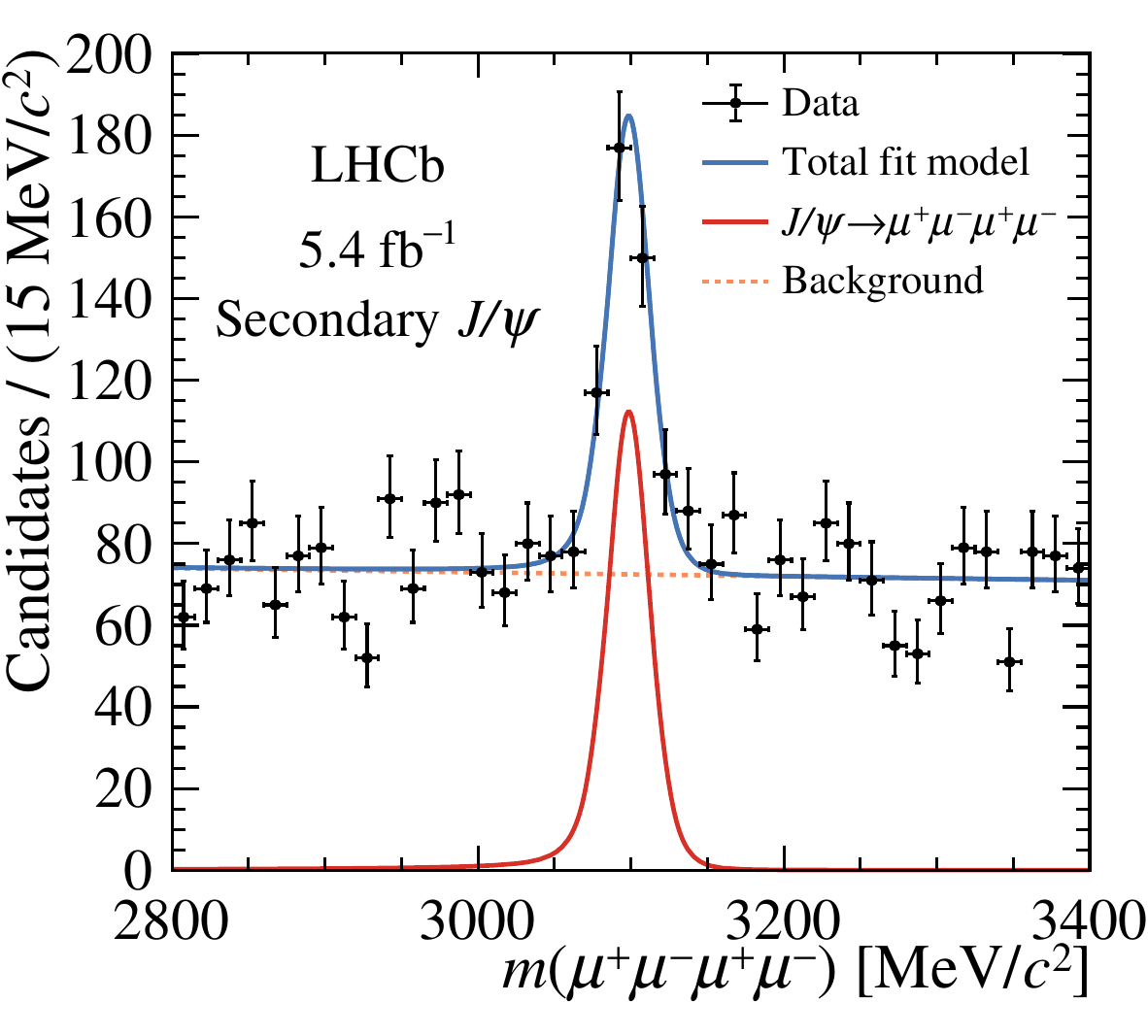}
    \end{subfigure}

    \caption{Invariant-mass distributions for (left) prompt $J\mskip -3mu/\mskip -2mu\psi \to \mu^+\mu^-\mu^+\mu^-$ candidates and (right) secondary $J\mskip -3mu/\mskip -2mu\psi \to \mu^+\mu^-\mu^+\mu^-$ candidates. The results of the fit are also shown.}
    \label{fig:Jpsi24mfit}
\end{figure}

Unbinned maximum-likelihood fits are used to describe the \jpsitofourm datasets. 
The signal shape is parameterised by a linear combination of two double-sided Crystal Ball functions sharing the same mean. As previously, the shape parameters are determined from simulation and are kept fixed when fitting to data, except for one mean and one resolution parameter that are allowed to vary, as done for the dimuon fit described above.  The background distribution is parameterised by an exponential function with a slope parameter, which is allowed to vary. 
In the prompt category, the \jpsi candidate is constrained to originate from the PV, which slightly improves the resolution on the four-muon invariant mass. 
The \jpsitofourm yield is expressed as a product of the ratio of branching fractions of \jpsitofourm and \jpsitomm decays ($R_{\mathcal{B}}$), the ratio of efficiencies between these two decay modes, and the previously measured \jpsitomm yield. The quantity $R_{\mathcal{B}}$ is therefore a parameter extracted directly from the fit. 
The resulting four-muon invariant-mass distributions, with the fits overlaid, are shown in Fig.~\ref{fig:Jpsi24mfit}. A clear peak of \jpsitofourm decays is observed in both datasets, with the significance markedly exceeding the discovery threshold of five standard deviations. 
The measured values of $R_{\mathcal{B}}$ are $(1.86\pm0.30)\times 10^{-5}$ (taking into account the fit uncertainty correction, discussed below) and $(1.90\pm0.20)\times 10^{-5}$ in the prompt and secondary categories, respectively, where the quoted uncertainty is statistical only. The two categories show a good agreement.

If the fit is reformulated with the \jpsitofourm yield as a free parameter, yields of $166\pm27$ and $286\pm 30$ decays are measured in the prompt and secondary categories. One can note that, unlike in the dimuon case, the yield of prompt \mbox{\jpsitofourm} decays is lower compared to the secondary decays, which is explained by the lower selection efficiency in the prompt category.  

Several cross-checks are performed to ensure the accuracy of the obtained results. 
The behaviour of each fit model is validated using a large number of pseudoexperiments. While good behaviour is found in the secondary category, the lower signal purity in the prompt category leads to an underestimation of the uncertainty by about 4\%. A correction of 4\% is therefore applied to the uncertainty returned by the fit in the prompt case.  
Fits are repeated in a narrower mass range, or by fixing the resolution parameter under the assumption that the resolution differs between the data and simulation by the same amount in the dimuon and four-muon datasets, and consistent results are obtained. 
The results are also determined separately for each data-taking period and are found to agree within the expected statistical scatter. 

\begin{figure}[!tb]
\centering
 
    \begin{subfigure}[t]{0.5\textwidth}
    \centering
    \includegraphics[height=2.3in]{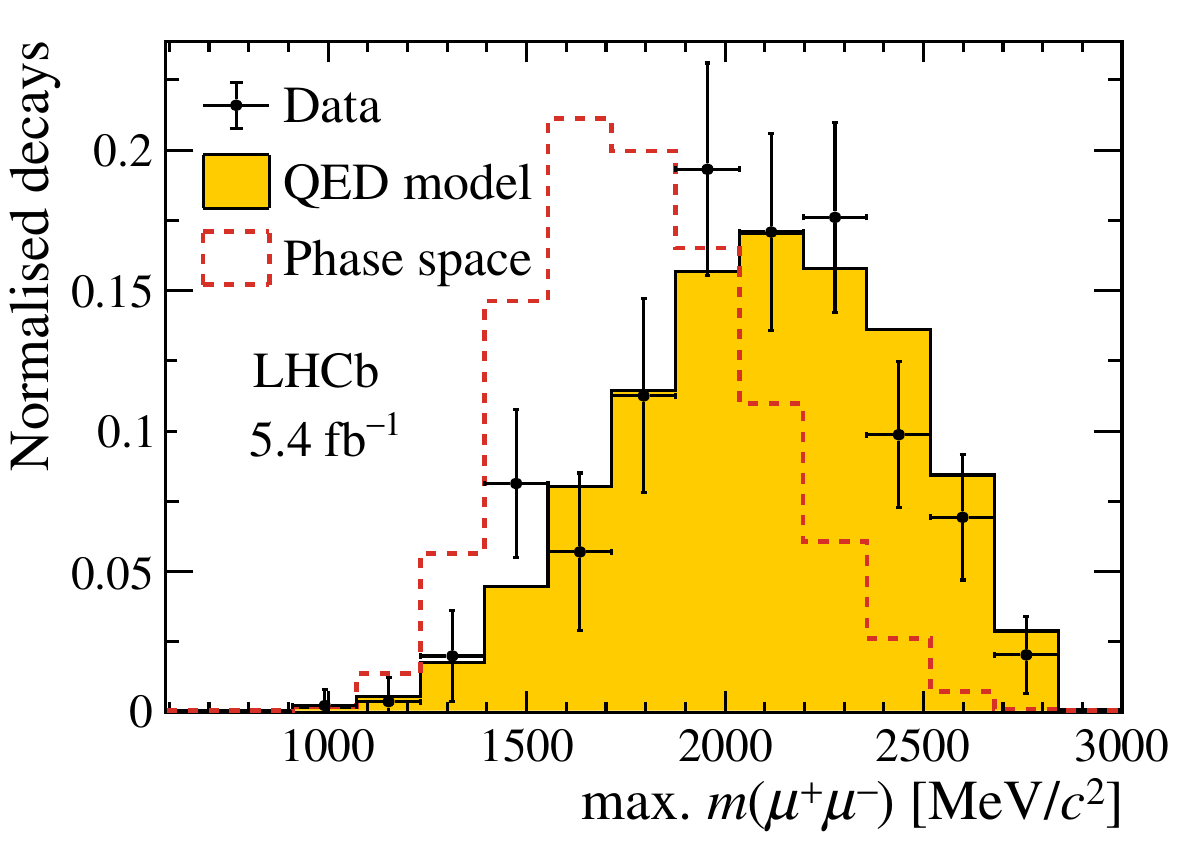}
    \end{subfigure}%
    \begin{subfigure}[t]{0.5\textwidth}
    \centering
    \includegraphics[height=2.3in]{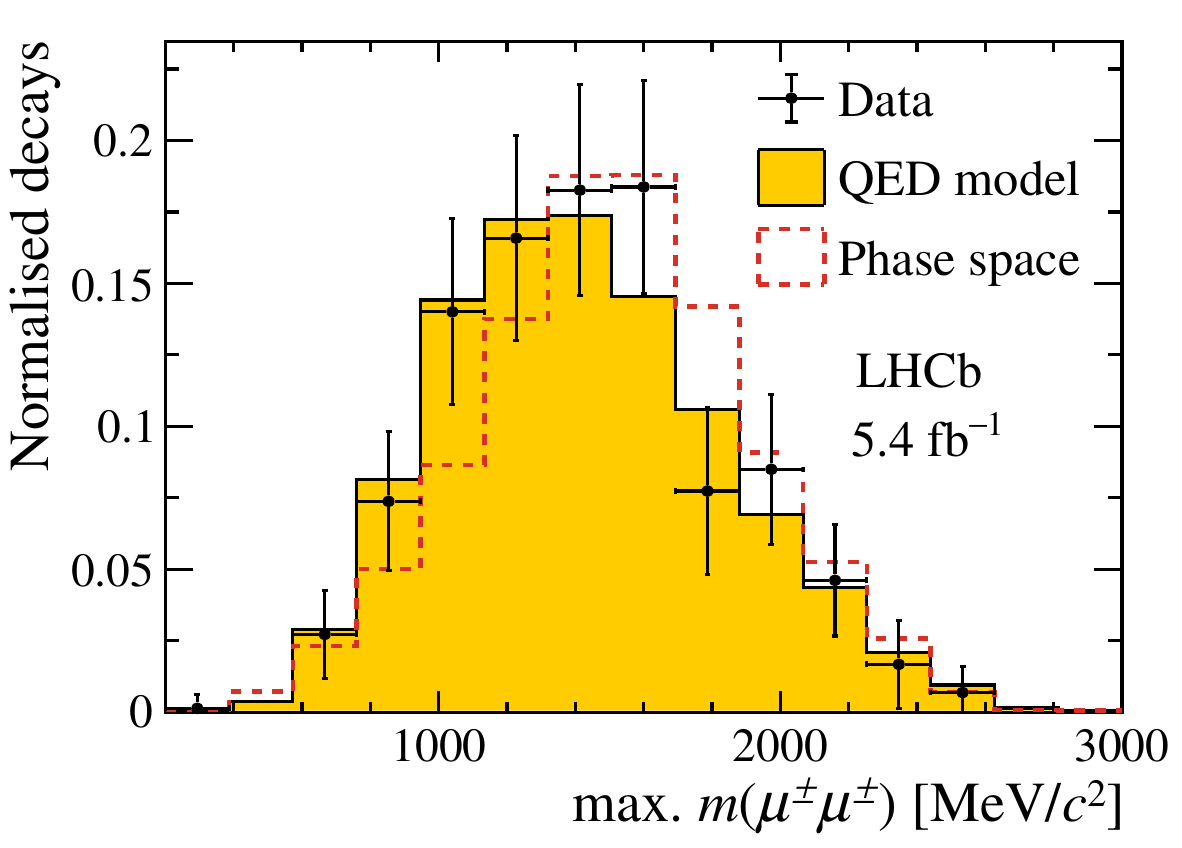}
    \end{subfigure}
    \begin{subfigure}[t]{0.5\textwidth}
    \centering
    \includegraphics[height=2.3in]{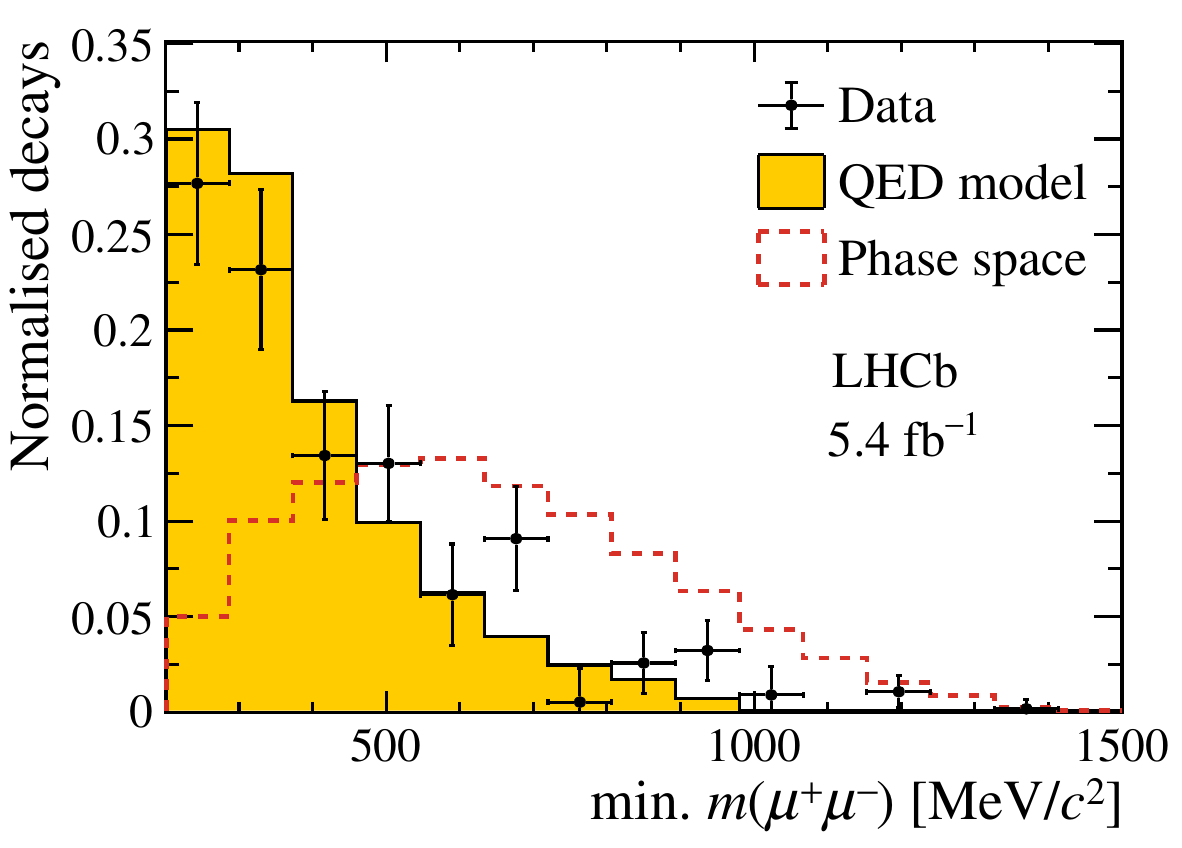}
    \end{subfigure}%
    \begin{subfigure}[t]{0.5\textwidth}
    \centering
    \includegraphics[height=2.3in]{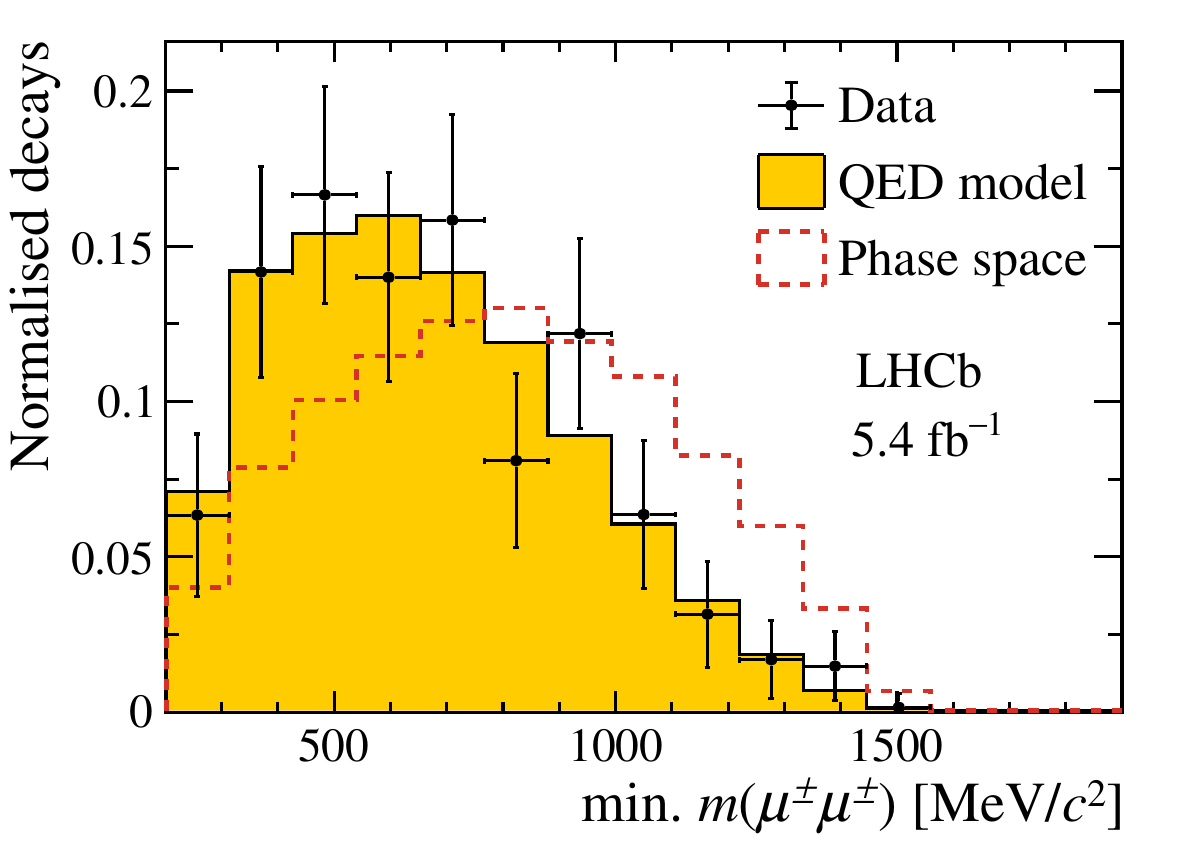}
    \end{subfigure}

    \caption{Comparison of maximum and minimum values of the invariant masses of the same-charge or opposite-charge muon pairs in background-subtracted data and calibrated simulation relying on the QED model in the secondary $J\mskip -3mu/\mskip -2mu\psi \to \mu^+\mu^-\mu^+\mu^-$ dataset. The distributions are normalised to the same area. The simulated phase-space model is added for comparison. For the data, only statistical uncertainties are shown.}
    \label{fig:dataMCagreement_4mu_det}
\end{figure}

Agreement between the background-subtracted data and calibrated simulation is examined in the kinematic observables of interest in the \jpsitofourm decays from the secondary category, as it offers the best statistical power. The background subtraction is performed using the \sPlot method~\cite{Pivk:2004ty}, using the four-muon invariant mass as a discriminating variable. 
As shown in Fig.~\ref{fig:dataMCagreement_4mu_det}, the dimuon invariant-mass distributions match well within statistical uncertainties, showing that the QED model describes the data accurately at the current precision level. No significant resonant structures are observed: this confirms that the decay is indeed dominated by the FSR process, as expected. 
In contrast, the model with kinematics uniformly distributed across the phase space does not match the data. 

As discussed above, the efficiency changes significantly throughout the phase space of the decay, especially in the prompt category. This could lead to biases if the data differ from the QED model at a level below the precision of the current data--model comparison (Fig.~\ref{fig:dataMCagreement_4mu_det}). 
Before studying this effect with two tests discussed below, the expected statistical scatter of efficiencies is estimated. 
This is achieved by generating datasets based on the QED model for \jpsitofourm decay, equal in statistical power to the signal observed in data. The standard deviation of the efficiencies determined on these datasets is found to be about 8\% (15\%) for the secondary (prompt) categories.  

The first test relies on performing a background subtraction of the \jpsitofourm signal and weighting the simulation to match the dimuon invariant-mass distributions in the data. Smoothing is applied to cope with the limited statistical power of the dataset. The resulting shift in efficiencies is about 6\% for both prompt and secondary categories, which is within the expected statistical scatter. 
The second test is based on parameterising the efficiency ratio in bins of three dimuon-mass observables, weighting the data by the efficiency correction factors and performing a weighted fit to extract the $R_{\mathcal{B}}$ parameter. This approach has several shortcomings: it assumes that the efficiency can be described as a function of only three observables; it is sensitive to the limitation in the number of simulated events; and it produces large correction weights in certain areas of the phase space, in particular, for background candidates. 
The obtained results are 4\% (15\%) lower compared to the baseline values for the secondary (prompt) categories, which is nevertheless consistent with the expected statistical scatter.

\section{Systematic uncertainties}
\label{sec:syst}

Several systematic effects, affecting both the efficiency estimation and the mass modelling, are studied. 

The uncertainty due to the limited size of the simulated sample is propagated to the final result. A small fraction of simulated signal candidates fails the truth-matching algorithm. Studies show that this fraction is larger for the four-muon dataset compared to the dimuon dataset and an uncertainty is assigned for the non cancellation in the ratio. 
Furthermore, the proportion of \jpsi mesons in simulation produced through the prompt production mechanism or from \bquark-hadron decays is altered within the uncertainty on its value from Ref.~\cite{LHCb-PAPER-2015-037}. 

For each correction applied to the signal simulation, a corresponding uncertainty is estimated.
The calibrations of the kinematics of the \jpsi meson, as well as the hardware trigger response, are re-evaluated with alternative binning schemes. An alternative dataset is also used to validate the tag-and-probe efficiency measurement for the hardware trigger. Similar studies are performed for the software trigger response. 
The uncertainties on the tracking efficiency calibration procedure are propagated to the final result. 

Detailed studies are performed to estimate the accuracy of the muon identification calibration. In the resampling procedure of the $P_\mu$ distributions, the kernel size is modified and the resampling is repeated. This allows the assessment of the uncertainty due to nonuniform statistical coverage of the calibration sample.
Furthermore, the bias coming from the inaccurate background subtraction procedure in the calibration samples is studied. This effect is pronounced at low muon \pt and therefore affects the four-muon datasets more than the dimuon ones. The associated systematic uncertainty is estimated by performing dedicated fits to the calibration data in intervals of \pt and measuring the difference in muon efficiency as compared to the background-subtracted data. 

Alternative models are studied to parameterise the signal and background shapes in the invariant-mass fits. For the signal shapes in each dataset, a Hypatia function~\cite{Santos:2013gra} is used. For the background shapes, a second-degree polynomial function is used in the four-muon dataset, with the parameters of the polynomial extracted from a fit to a wider mass range. In the dimuon datasets, an exponential shape is used. 
Each alternative model is used to generate a large number of pseudodatasets, and the fit with the default model is repeated to estimate the average bias. As expected, in the dimuon datasets this systematic uncertainty is larger compared to the statistical uncertainty, however, it is still negligible compared to the statistical precision of the \jpsitofourm yield. 

The procedure to remove multiple candidates is also scrutinised. 
In the dimuon datasets, where all candidates are preserved in the default approach, only one arbitrarily chosen candidate is kept. 
In the secondary four-muon dataset, where the fraction of the multiple candidates is about 1\% and only one candidate is retained in the default approach, all candidates are kept.
Finally, in the prompt four-muon dataset, where the reconstruction procedure leads to the presence of duplicate candidates for most events, it is not meaningful to retain all candidates. Instead, the criterion for selecting the one candidate to be kept is modified by selecting the candidates that are rejected by the default procedure.

A summary of the systematic uncertainties on $R_{\mathcal{B}}$ is presented in Table~\ref{tab:syst}. Each source of uncertainty is assumed to be either fully correlated or fully uncorrelated between the two analysis categories. 

\begin{table}[tb!]
\centering
\caption{Summary of systematic uncertainties considered in this analysis. All values are quoted in percent of $R_{\mathcal{B}}$. The relative statistical uncertainty is also quoted for reference.}
\label{tab:syst}

\begin{tabular}{l c c c}
Source & Secondary & Prompt & Correlated \\
\hline 
Size of simulated samples & 0.6 & 0.8 & no\\
Treatment of simulation  & 1.1 & 0.8 & yes \\
Fraction of \jpsi component from \bquark decays & 0.3 & 0.2 & no \\
Kinematic calibration  & 0.5 & 0.2 & no \\
Hardware trigger calibration & 0.3 &0.4 & yes\\
Software trigger calibration & 1.0 & 0.4 & no\\
Tracking calibration & 2.3 & 1.8 & yes\\
Muon ID calibration  & 3.1 & 5.8 & yes\\
Mass modelling & 0.6 & 0.7 & no \\
Multiple candidate selection & 0.6 & 0.8 & no\\
\hline
Total systematic uncertainty & 4.4 & 6.3 & --\\
\hline
Statistical uncertainty & 10.5& 16.2& no\\
\end{tabular}
\end{table}

Finally, the known value of $\mathcal{B}(\jpsitomm) = (5.961\pm0.033)\%$~\cite{PDG2024} is used to determine the absolute value of $\mathcal{B}(\jpsitofourm)$. The uncertainty on this value is propagated to the final result as a separate term. 

\section{Results and conclusion}
\label{sec:Conclusion}

In the secondary category of \jpsi decays, the ratio of branching fractions of the \mbox{\jpsitofourm} and \jpsitomm decays is found to be 
$    R_{\mathcal{B}} = (1.90\pm0.20\pm0.08)\times 10^{-5}$,
where the first uncertainty is statistical and second systematic.
In the prompt category of \jpsi decays, the measured value is $R_{\mathcal{B}} = (1.86\pm0.30\pm0.12)\times 10^{-5}$. These two results are in excellent agreement and show stability with respect to splits by data-taking periods.

The results from the two analysis categories are combined by performing a weighted average that provides the minimum variance estimate of $R_{\mathcal{B}}$. Statistical uncertainties are treated as uncorrelated. Systematic uncertainties are treated as either fully correlated or uncorrelated, as described in Table~\ref{tab:syst}, for which their total correlation coefficient is 0.10. 
This results in the value
\begin{equation*}
    R_{\mathcal{B}} = (1.89\pm0.17\pm0.09)\times 10^{-5},
\end{equation*}
which agrees within 1.4 standard deviations with the leading-order QED calculation of ${R_{\mathcal{B}}=1.63\times10^{-5}}$ from Ref.~\cite{Chen:2020bju}.

\begin{figure}[!b]
        \centering
    \includegraphics[width=0.8\textwidth]{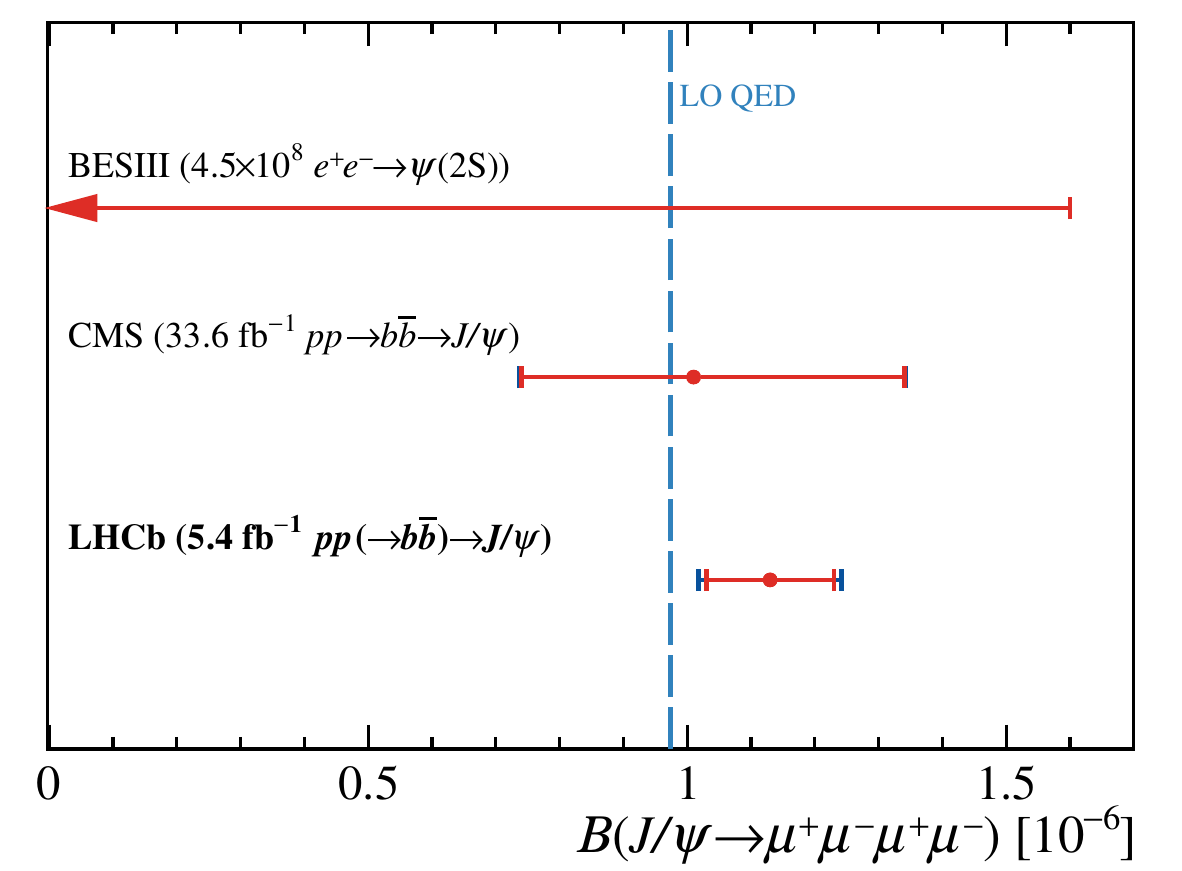}

    \caption{Summary of the measurements of the $J\mskip -3mu/\mskip -2mu\psi \to \mu^+\mu^-\mu^+\mu^-$ branching fraction by the \mbox{BESIII}~\cite{BESIII:2021ocn} (upper limit at 90\% CL) and CMS~\cite{CMS:2024jpsi} experiments, compared with the result from this paper. The statistical uncertainties are shown in red, with the total uncertainties in blue. The leading-order QED prediction from Ref.~\cite{Chen:2020bju} is shown as a dashed line whose uncertainty is too small to display.}
    \label{fig:summary}
\end{figure}

Multiplying the measured value of $R_{\mathcal{B}}$ by the known value of the dimuon branching fraction of the \jpsi meson~\cite{PDG2024}, 
\begin{equation*}
   \mathcal{B}(\jpsitofourm) = (1.13\pm0.10\pm0.05\pm0.01)\times 10^{-6},
\end{equation*}
is obtained, where the first uncertainty is statistical, the second is systematic and the third is due to the uncertainty on $\mathcal{B}(\jpsitomm)$. This result relies on the simulated decay model based on the QED calculation. 
Indeed, the dimuon invariant-mass distributions in the background-subtracted \jpsitofourm data agree, within available statistical precision, with the expectations from the QED calculation; no significant resonant structures are observed. A summary of existing measurements of the \jpsitofourm branching fraction is shown in Fig.~\ref{fig:summary}.

To conclude, this analysis presents the first measurement of a four-lepton decay of quarkonium at the LHCb experiment. 
The \jpsitofourm decay is observed with significance far above the discovery threshold, and the most precise measurement to date of its branching fraction is performed. 
This result provides complementary input to the study of $\jpsi \to \epem\epem$ and $\jpsi \to \epem \mumu$ decays by the \besiii  experiment~\cite{BESIII:2021ocn}. It also is consistent with the recent measurement by the CMS collaboration~\cite{CMS:2024jpsi}.  
To validate the robustness of the analysis procedure, the data are divided into two disjoint samples, corresponding to prompt and secondary \jpsi meson production, and analysed separately. 
The \jpsitomm decay is used as a normalisation mode, also split by the production mechanism. 
The kinematic distributions in the \jpsitofourm decay are found to agree with the QED model, but differ significantly from the phase-space model. 

\enlargethispage{\baselineskip}
This observation of the \jpsitofourm decay paves the way for studies of other multilepton decays at LHCb. 
The importance of correct decay-model parametrisation in simulation is highlighted: the impact of the lepton kinematic selection  on the decay-model dependence should be taken into account in these future analyses.

\section*{Acknowledgements}
%
%
\noindent We thank the \besiii collaboration and Jianping Dai for providing us with the \evtgen model for the decay of interest. We express our gratitude to our colleagues in the CERN
accelerator departments for the excellent performance of the LHC. We
thank the technical and administrative staff at the LHCb
institutes.
We acknowledge support from CERN and from the national agencies:
CAPES, CNPq, FAPERJ and FINEP (Brazil); 
MOST and NSFC (China); 
CNRS/IN2P3 (France); 
BMBF, DFG and MPG (Germany); 
INFN (Italy); 
NWO (Netherlands); 
MNiSW and NCN (Poland); 
MCID/IFA (Romania); 
MICIU and AEI (Spain);
SNSF and SER (Switzerland); 
NASU (Ukraine); 
STFC (United Kingdom); 
DOE NP and NSF (USA).
We acknowledge the computing resources that are provided by CERN, IN2P3
(France), KIT and DESY (Germany), INFN (Italy), SURF (Netherlands),
PIC (Spain), GridPP (United Kingdom), 
CSCS (Switzerland), IFIN-HH (Romania), CBPF (Brazil),
and Polish WLCG (Poland).
We are indebted to the communities behind the multiple open-source
software packages on which we depend.
Individual groups or members have received support from
ARC and ARDC (Australia);
Key Research Program of Frontier Sciences of CAS, CAS PIFI, CAS CCEPP, 
Fundamental Research Funds for the Central Universities, 
and Sci. \& Tech. Program of Guangzhou (China);
Minciencias (Colombia);
EPLANET, Marie Sk\l{}odowska-Curie Actions, ERC and NextGenerationEU (European Union);
A*MIDEX, ANR, IPhU and Labex P2IO, and R\'{e}gion Auvergne-Rh\^{o}ne-Alpes (France);
AvH Foundation (Germany);
ICSC (Italy); 
Severo Ochoa and Mar\'ia de Maeztu Units of Excellence, GVA, XuntaGal, GENCAT, InTalent-Inditex and Prog. ~Atracci\'on Talento CM (Spain);
SRC (Sweden);
the Leverhulme Trust, the Royal Society
 and UKRI (United Kingdom).

\clearpage

\addcontentsline{toc}{section}{References}
\bibliographystyle{LHCb}
\bibliography{main,standard,LHCb-PAPER,LHCb-CONF,LHCb-DP,LHCb-TDR}

\ifx\mcitethebibliography\mciteundefinedmacro
\PackageError{LHCb.bst}{mciteplus.sty has not been loaded}
{This bibstyle requires the use of the mciteplus package.}\fi
\providecommand{\href}[2]{#2}
\begin{mcitethebibliography}{10}
\mciteSetBstSublistMode{n}
\mciteSetBstMaxWidthForm{subitem}{\alph{mcitesubitemcount})}
\mciteSetBstSublistLabelBeginEnd{\mcitemaxwidthsubitemform\space}
{\relax}{\relax}

\bibitem{Chala:2019vzu}
M.~Chala, U.~Egede, and M.~Spannowsky, \ifthenelse{\boolean{articletitles}}{\emph{{Searching new physics in rare $B$-meson decays into multiple muons}}, }{}\href{https://doi.org/10.1140/epjc/s10052-019-6946-6}{Eur.\ Phys.\ J.\ C \textbf{79} (2019) 431}, \href{http://arxiv.org/abs/1902.10156}{{\normalfont\ttfamily arXiv:1902.10156}}\relax
\mciteBstWouldAddEndPuncttrue
\mciteSetBstMidEndSepPunct{\mcitedefaultmidpunct}
{\mcitedefaultendpunct}{\mcitedefaultseppunct}\relax
\EndOfBibitem
\bibitem{PDG2024}
Particle Data Group, S.~Navas {\em et~al.}, \ifthenelse{\boolean{articletitles}}{\emph{{\href{http://pdg.lbl.gov/}{Review of particle physics}}}, }{}\href{https://doi.org/10.1103/PhysRevD.110.030001}{Phys.\ Rev.\  \textbf{D110} (2024) 030001}\relax
\mciteBstWouldAddEndPuncttrue
\mciteSetBstMidEndSepPunct{\mcitedefaultmidpunct}
{\mcitedefaultendpunct}{\mcitedefaultseppunct}\relax
\EndOfBibitem
\bibitem{BESIII:2022cul}
BESIII collaboration, M.~Ablikim {\em et~al.}, \ifthenelse{\boolean{articletitles}}{\emph{{Observation of the double Dalitz decay $\eta'\to e^+e^-e^+e^-$}}, }{}\href{https://doi.org/10.1103/PhysRevD.105.112010}{Phys.\ Rev.\  \textbf{D105} (2022) 112010}, \href{http://arxiv.org/abs/2203.12229}{{\normalfont\ttfamily arXiv:2203.12229}}\relax
\mciteBstWouldAddEndPuncttrue
\mciteSetBstMidEndSepPunct{\mcitedefaultmidpunct}
{\mcitedefaultendpunct}{\mcitedefaultseppunct}\relax
\EndOfBibitem
\bibitem{CMS:2023gui}
CMS collaboration, A.~Hayrapetyan {\em et~al.}, \ifthenelse{\boolean{articletitles}}{\emph{{Observation of the rare decay of the $\eta$ meson to four muons}}, }{}\href{https://doi.org/10.1103/PhysRevLett.131.091903}{Phys.\ Rev.\ Lett.\  \textbf{131} (2023) 091903}, \href{http://arxiv.org/abs/2305.04904}{{\normalfont\ttfamily arXiv:2305.04904}}\relax
\mciteBstWouldAddEndPuncttrue
\mciteSetBstMidEndSepPunct{\mcitedefaultmidpunct}
{\mcitedefaultendpunct}{\mcitedefaultseppunct}\relax
\EndOfBibitem
\bibitem{LHCb-PAPER-2022-035}
LHCb collaboration, R.~Aaij {\em et~al.}, \ifthenelse{\boolean{articletitles}}{\emph{{Search for $K^0_{S(L)} \to \mup\mun\mup\mun$ decays at LHCb}}, }{}\href{https://doi.org/10.1103/PhysRevD.108.L031102}{Phys.\ Rev.\  \textbf{D108} (2023) L031102}, \href{http://arxiv.org/abs/2212.04977}{{\normalfont\ttfamily arXiv:2212.04977}}\relax
\mciteBstWouldAddEndPuncttrue
\mciteSetBstMidEndSepPunct{\mcitedefaultmidpunct}
{\mcitedefaultendpunct}{\mcitedefaultseppunct}\relax
\EndOfBibitem
\bibitem{LHCb-PAPER-2021-039}
LHCb collaboration, R.~Aaij {\em et~al.}, \ifthenelse{\boolean{articletitles}}{\emph{{Search for rare $B^0_{(s)} \to \mu^+ \mu^- \mu^+ \mu^-$ decays}}, }{}\href{https://doi.org/10.1007/JHEP03(2022)109}{JHEP \textbf{03} (2022) 109}, \href{http://arxiv.org/abs/2111.11339}{{\normalfont\ttfamily arXiv:2111.11339}}\relax
\mciteBstWouldAddEndPuncttrue
\mciteSetBstMidEndSepPunct{\mcitedefaultmidpunct}
{\mcitedefaultendpunct}{\mcitedefaultseppunct}\relax
\EndOfBibitem
\bibitem{Danilina:2018uzr}
A.~V. Danilina and N.~V. Nikitin, \ifthenelse{\boolean{articletitles}}{\emph{{Four-leptonic decays of charged and neutral $B$ mesons within the Standard Model}}, }{}\href{https://doi.org/10.1134/S1063778818030092}{Phys.\ Atom.\ Nucl.\  \textbf{81} (2018) 347}\relax
\mciteBstWouldAddEndPuncttrue
\mciteSetBstMidEndSepPunct{\mcitedefaultmidpunct}
{\mcitedefaultendpunct}{\mcitedefaultseppunct}\relax
\EndOfBibitem
\bibitem{Chen:2020bju}
W.~Chen {\em et~al.}, \ifthenelse{\boolean{articletitles}}{\emph{{Four-lepton decays of neutral vector mesons}}, }{}\href{https://doi.org/10.1103/PhysRevD.104.094023}{Phys.\ Rev.\  \textbf{D104} (2021) 094023}, \href{http://arxiv.org/abs/2009.12363}{{\normalfont\ttfamily arXiv:2009.12363}}\relax
\mciteBstWouldAddEndPuncttrue
\mciteSetBstMidEndSepPunct{\mcitedefaultmidpunct}
{\mcitedefaultendpunct}{\mcitedefaultseppunct}\relax
\EndOfBibitem
\bibitem{Kopke:1988cs}
L.~Kopke and N.~Wermes, \ifthenelse{\boolean{articletitles}}{\emph{{$J/\psi$ decays}}, }{}\href{https://doi.org/10.1016/0370-1573(89)90074-4}{Phys.\ Rept.\  \textbf{174} (1989) 67}\relax
\mciteBstWouldAddEndPuncttrue
\mciteSetBstMidEndSepPunct{\mcitedefaultmidpunct}
{\mcitedefaultendpunct}{\mcitedefaultseppunct}\relax
\EndOfBibitem
\bibitem{Voloshin:2007dx}
M.~B. Voloshin, \ifthenelse{\boolean{articletitles}}{\emph{{Charmonium}}, }{}\href{https://doi.org/10.1016/j.ppnp.2008.02.001}{Prog.\ Part.\ Nucl.\ Phys.\  \textbf{61} (2008) 455}, \href{http://arxiv.org/abs/0711.4556}{{\normalfont\ttfamily arXiv:0711.4556}}\relax
\mciteBstWouldAddEndPuncttrue
\mciteSetBstMidEndSepPunct{\mcitedefaultmidpunct}
{\mcitedefaultendpunct}{\mcitedefaultseppunct}\relax
\EndOfBibitem
\bibitem{BESIII:2021ocn}
BESIII collaboration, M.~Ablikim {\em et~al.}, \ifthenelse{\boolean{articletitles}}{\emph{{Observation of \jpsi decays to $\epem\epem$ and $\epem\mumu$}}, }{}\href{https://doi.org/10.1103/PhysRevD.109.052006}{Phys.\ Rev.\  \textbf{D109} (2024) 052006}, \href{http://arxiv.org/abs/2111.13881}{{\normalfont\ttfamily arXiv:2111.13881}}\relax
\mciteBstWouldAddEndPuncttrue
\mciteSetBstMidEndSepPunct{\mcitedefaultmidpunct}
{\mcitedefaultendpunct}{\mcitedefaultseppunct}\relax
\EndOfBibitem
\bibitem{CMS:2024jpsi}
CMS collaboration, A.~Hayrapetyan {\em et~al.}, \ifthenelse{\boolean{articletitles}}{\emph{{Observation of the $\jpsi \to \mumu\mumu$ decay in proton-proton collisions at s=13\,\,TeV}}, }{}\href{https://doi.org/10.1103/PhysRevD.109.L111101}{Phys.\ Rev.\  \textbf{D109} (2024) L111101}, \href{http://arxiv.org/abs/2403.11352}{{\normalfont\ttfamily arXiv:2403.11352}}\relax
\mciteBstWouldAddEndPuncttrue
\mciteSetBstMidEndSepPunct{\mcitedefaultmidpunct}
{\mcitedefaultendpunct}{\mcitedefaultseppunct}\relax
\EndOfBibitem
\bibitem{LHCb-PAPER-2015-037}
LHCb collaboration, R.~Aaij {\em et~al.}, \ifthenelse{\boolean{articletitles}}{\emph{{Measurement of forward \jpsi production cross-sections in \proton\proton collisions at \mbox{$\sqs=$13~\tev}}}, }{}\href{https://doi.org/10.1007/JHEP10(2015)172}{JHEP \textbf{10} (2015) 172}, Erratum \href{https://doi.org/10.1007/JHEP05(2017)063}{ibid.\   \textbf{05} (2017) 063}, \href{http://arxiv.org/abs/1509.00771}{{\normalfont\ttfamily arXiv:1509.00771}}\relax
\mciteBstWouldAddEndPuncttrue
\mciteSetBstMidEndSepPunct{\mcitedefaultmidpunct}
{\mcitedefaultendpunct}{\mcitedefaultseppunct}\relax
\EndOfBibitem
\bibitem{LHCb-PAPER-2018-002}
LHCb collaboration, R.~Aaij {\em et~al.}, \ifthenelse{\boolean{articletitles}}{\emph{{Measurement of \Upsilonres production cross-section in \proton\proton collisions at $\sqs = $13~\tev}}, }{}\href{https://doi.org/10.1007/JHEP07(2018)134}{JHEP \textbf{07} (2018) 134}, \href{http://arxiv.org/abs/1804.09214}{{\normalfont\ttfamily arXiv:1804.09214}}\relax
\mciteBstWouldAddEndPuncttrue
\mciteSetBstMidEndSepPunct{\mcitedefaultmidpunct}
{\mcitedefaultendpunct}{\mcitedefaultseppunct}\relax
\EndOfBibitem
\bibitem{LHCb-DP-2008-001}
LHCb collaboration, A.~A. Alves~Jr.\ {\em et~al.}, \ifthenelse{\boolean{articletitles}}{\emph{{The \lhcb detector at the LHC}}, }{}\href{https://doi.org/10.1088/1748-0221/3/08/S08005}{JINST \textbf{3} (2008) S08005}\relax
\mciteBstWouldAddEndPuncttrue
\mciteSetBstMidEndSepPunct{\mcitedefaultmidpunct}
{\mcitedefaultendpunct}{\mcitedefaultseppunct}\relax
\EndOfBibitem
\bibitem{LHCb-DP-2014-002}
LHCb collaboration, R.~Aaij {\em et~al.}, \ifthenelse{\boolean{articletitles}}{\emph{{LHCb detector performance}}, }{}\href{https://doi.org/10.1142/S0217751X15300227}{Int.\ J.\ Mod.\ Phys.\  \textbf{A30} (2015) 1530022}, \href{http://arxiv.org/abs/1412.6352}{{\normalfont\ttfamily arXiv:1412.6352}}\relax
\mciteBstWouldAddEndPuncttrue
\mciteSetBstMidEndSepPunct{\mcitedefaultmidpunct}
{\mcitedefaultendpunct}{\mcitedefaultseppunct}\relax
\EndOfBibitem
\bibitem{LHCb-DP-2013-001}
F.~Archilli {\em et~al.}, \ifthenelse{\boolean{articletitles}}{\emph{{Performance of the muon identification at LHCb}}, }{}\href{https://doi.org/10.1088/1748-0221/8/10/P10020}{JINST \textbf{8} (2013) P10020}, \href{http://arxiv.org/abs/1306.0249}{{\normalfont\ttfamily arXiv:1306.0249}}\relax
\mciteBstWouldAddEndPuncttrue
\mciteSetBstMidEndSepPunct{\mcitedefaultmidpunct}
{\mcitedefaultendpunct}{\mcitedefaultseppunct}\relax
\EndOfBibitem
\bibitem{LHCb-DP-2018-001}
R.~Aaij {\em et~al.}, \ifthenelse{\boolean{articletitles}}{\emph{{Selection and processing of calibration samples to measure the particle identification performance of the LHCb experiment in Run 2}}, }{}\href{https://doi.org/10.1140/epjti/s40485-019-0050-z}{Eur.\ Phys.\ J.\ Tech.\ Instr.\  \textbf{6} (2019) 1}, \href{http://arxiv.org/abs/1803.00824}{{\normalfont\ttfamily arXiv:1803.00824}}\relax
\mciteBstWouldAddEndPuncttrue
\mciteSetBstMidEndSepPunct{\mcitedefaultmidpunct}
{\mcitedefaultendpunct}{\mcitedefaultseppunct}\relax
\EndOfBibitem
\bibitem{LHCb-DP-2012-004}
R.~Aaij {\em et~al.}, \ifthenelse{\boolean{articletitles}}{\emph{{The \lhcb trigger and its performance in 2011}}, }{}\href{https://doi.org/10.1088/1748-0221/8/04/P04022}{JINST \textbf{8} (2013) P04022}, \href{http://arxiv.org/abs/1211.3055}{{\normalfont\ttfamily arXiv:1211.3055}}\relax
\mciteBstWouldAddEndPuncttrue
\mciteSetBstMidEndSepPunct{\mcitedefaultmidpunct}
{\mcitedefaultendpunct}{\mcitedefaultseppunct}\relax
\EndOfBibitem
\bibitem{Sjostrand:2007gs}
T.~Sj\"{o}strand, S.~Mrenna, and P.~Skands, \ifthenelse{\boolean{articletitles}}{\emph{{A brief introduction to PYTHIA 8.1}}, }{}\href{https://doi.org/10.1016/j.cpc.2008.01.036}{Comput.\ Phys.\ Commun.\  \textbf{178} (2008) 852}, \href{http://arxiv.org/abs/0710.3820}{{\normalfont\ttfamily arXiv:0710.3820}}\relax
\mciteBstWouldAddEndPuncttrue
\mciteSetBstMidEndSepPunct{\mcitedefaultmidpunct}
{\mcitedefaultendpunct}{\mcitedefaultseppunct}\relax
\EndOfBibitem
\bibitem{Sjostrand:2006za}
T.~Sj\"{o}strand, S.~Mrenna, and P.~Skands, \ifthenelse{\boolean{articletitles}}{\emph{{PYTHIA 6.4 physics and manual}}, }{}\href{https://doi.org/10.1088/1126-6708/2006/05/026}{JHEP \textbf{05} (2006) 026}, \href{http://arxiv.org/abs/hep-ph/0603175}{{\normalfont\ttfamily arXiv:hep-ph/0603175}}\relax
\mciteBstWouldAddEndPuncttrue
\mciteSetBstMidEndSepPunct{\mcitedefaultmidpunct}
{\mcitedefaultendpunct}{\mcitedefaultseppunct}\relax
\EndOfBibitem
\bibitem{LHCb-PROC-2010-056}
I.~Belyaev {\em et~al.}, \ifthenelse{\boolean{articletitles}}{\emph{{Handling of the generation of primary events in Gauss, the LHCb simulation framework}}, }{}\href{https://doi.org/10.1088/1742-6596/331/3/032047}{J.\ Phys.\ Conf.\ Ser.\  \textbf{331} (2011) 032047}\relax
\mciteBstWouldAddEndPuncttrue
\mciteSetBstMidEndSepPunct{\mcitedefaultmidpunct}
{\mcitedefaultendpunct}{\mcitedefaultseppunct}\relax
\EndOfBibitem
\bibitem{Lange:2001uf}
D.~J. Lange, \ifthenelse{\boolean{articletitles}}{\emph{{The EvtGen particle decay simulation package}}, }{}\href{https://doi.org/10.1016/S0168-9002(01)00089-4}{Nucl.\ Instrum.\ Meth.\  \textbf{A462} (2001) 152}\relax
\mciteBstWouldAddEndPuncttrue
\mciteSetBstMidEndSepPunct{\mcitedefaultmidpunct}
{\mcitedefaultendpunct}{\mcitedefaultseppunct}\relax
\EndOfBibitem
\bibitem{davidson2015photos}
N.~Davidson, T.~Przedzinski, and Z.~Was, \ifthenelse{\boolean{articletitles}}{\emph{{PHOTOS interface in C++: Technical and physics documentation}}, }{}\href{https://doi.org/https://doi.org/10.1016/j.cpc.2015.09.013}{Comp.\ Phys.\ Comm.\  \textbf{199} (2016) 86}, \href{http://arxiv.org/abs/1011.0937}{{\normalfont\ttfamily arXiv:1011.0937}}\relax
\mciteBstWouldAddEndPuncttrue
\mciteSetBstMidEndSepPunct{\mcitedefaultmidpunct}
{\mcitedefaultendpunct}{\mcitedefaultseppunct}\relax
\EndOfBibitem
\bibitem{Allison:2006ve}
Geant4 collaboration, J.~Allison {\em et~al.}, \ifthenelse{\boolean{articletitles}}{\emph{{Geant4 developments and applications}}, }{}\href{https://doi.org/10.1109/TNS.2006.869826}{IEEE Trans.\ Nucl.\ Sci.\  \textbf{53} (2006) 270}\relax
\mciteBstWouldAddEndPuncttrue
\mciteSetBstMidEndSepPunct{\mcitedefaultmidpunct}
{\mcitedefaultendpunct}{\mcitedefaultseppunct}\relax
\EndOfBibitem
\bibitem{Agostinelli:2002hh}
Geant4 collaboration, S.~Agostinelli {\em et~al.}, \ifthenelse{\boolean{articletitles}}{\emph{{Geant4: A simulation toolkit}}, }{}\href{https://doi.org/10.1016/S0168-9002(03)01368-8}{Nucl.\ Instrum.\ Meth.\  \textbf{A506} (2003) 250}\relax
\mciteBstWouldAddEndPuncttrue
\mciteSetBstMidEndSepPunct{\mcitedefaultmidpunct}
{\mcitedefaultendpunct}{\mcitedefaultseppunct}\relax
\EndOfBibitem
\bibitem{LHCb-PROC-2011-006}
M.~Clemencic {\em et~al.}, \ifthenelse{\boolean{articletitles}}{\emph{{The \lhcb simulation application, Gauss: Design, evolution and experience}}, }{}\href{https://doi.org/10.1088/1742-6596/331/3/032023}{J.\ Phys.\ Conf.\ Ser.\  \textbf{331} (2011) 032023}\relax
\mciteBstWouldAddEndPuncttrue
\mciteSetBstMidEndSepPunct{\mcitedefaultmidpunct}
{\mcitedefaultendpunct}{\mcitedefaultseppunct}\relax
\EndOfBibitem
\bibitem{LHCb-DP-2013-002}
LHCb collaboration, R.~Aaij {\em et~al.}, \ifthenelse{\boolean{articletitles}}{\emph{{Measurement of the track reconstruction efficiency at LHCb}}, }{}\href{https://doi.org/10.1088/1748-0221/10/02/P02007}{JINST \textbf{10} (2015) P02007}, \href{http://arxiv.org/abs/1408.1251}{{\normalfont\ttfamily arXiv:1408.1251}}\relax
\mciteBstWouldAddEndPuncttrue
\mciteSetBstMidEndSepPunct{\mcitedefaultmidpunct}
{\mcitedefaultendpunct}{\mcitedefaultseppunct}\relax
\EndOfBibitem
\bibitem{LHCb-PUB-2014-039}
S.~Tolk, J.~Albrecht, F.~Dettori, and A.~Pellegrino, \ifthenelse{\boolean{articletitles}}{\emph{{Data driven trigger efficiency determination at LHCb}}, }{} \href{http://cdsweb.cern.ch/search?p=LHCb-PUB-2014-039&f=reportnumber&action_search=Search&c=LHCb+Notes} {LHCb-PUB-2014-039}, 2014\relax
\mciteBstWouldAddEndPuncttrue
\mciteSetBstMidEndSepPunct{\mcitedefaultmidpunct}
{\mcitedefaultendpunct}{\mcitedefaultseppunct}\relax
\EndOfBibitem
\bibitem{LHCb-DP-2019-001}
R.~Aaij {\em et~al.}, \ifthenelse{\boolean{articletitles}}{\emph{{Design and performance of the LHCb trigger and full real-time reconstruction in Run 2 of the LHC}}, }{}\href{https://doi.org/10.1088/1748-0221/14/04/P04013}{JINST \textbf{14} (2019) P04013}, \href{http://arxiv.org/abs/1812.10790}{{\normalfont\ttfamily arXiv:1812.10790}}\relax
\mciteBstWouldAddEndPuncttrue
\mciteSetBstMidEndSepPunct{\mcitedefaultmidpunct}
{\mcitedefaultendpunct}{\mcitedefaultseppunct}\relax
\EndOfBibitem
\bibitem{LHCb-PAPER-2023-022}
LHCb collaboration, R.~Aaij {\em et~al.}, \ifthenelse{\boolean{articletitles}}{\emph{{Measurement of $\jpsi$-pair production in $\proton\proton$ collisions at $\sqrt{s} = 13$~\tev and study of gluon transverse-momentum dependent PDFs}}, }{}\href{https://doi.org/10.1007/JHEP03(2024)088}{JHEP \textbf{03} (2024) 088}, \href{http://arxiv.org/abs/2311.14085}{{\normalfont\ttfamily arXiv:2311.14085}}\relax
\mciteBstWouldAddEndPuncttrue
\mciteSetBstMidEndSepPunct{\mcitedefaultmidpunct}
{\mcitedefaultendpunct}{\mcitedefaultseppunct}\relax
\EndOfBibitem
\bibitem{Breiman}
L.~Breiman, J.~H. Friedman, R.~A. Olshen, and C.~J. Stone, {\em Classification and regression trees}, Wadsworth international group, Belmont, California, USA, 1984\relax
\mciteBstWouldAddEndPuncttrue
\mciteSetBstMidEndSepPunct{\mcitedefaultmidpunct}
{\mcitedefaultendpunct}{\mcitedefaultseppunct}\relax
\EndOfBibitem
\bibitem{AdaBoost}
Y.~Freund and R.~E. Schapire, \ifthenelse{\boolean{articletitles}}{\emph{A decision-theoretic generalization of on-line learning and an application to boosting}, }{}\href{https://doi.org/10.1006/jcss.1997.1504}{J.\ Comput.\ Syst.\ Sci.\  \textbf{55} (1997) 119}\relax
\mciteBstWouldAddEndPuncttrue
\mciteSetBstMidEndSepPunct{\mcitedefaultmidpunct}
{\mcitedefaultendpunct}{\mcitedefaultseppunct}\relax
\EndOfBibitem
\bibitem{Hocker:2007ht}
H.~Voss, A.~Hoecker, J.~Stelzer, and F.~Tegenfeldt, \ifthenelse{\boolean{articletitles}}{\emph{{TMVA - Toolkit for Multivariate Data Analysis with ROOT}}, }{}\href{https://doi.org/10.22323/1.050.0040}{PoS \textbf{ACAT} (2007) 040}\relax
\mciteBstWouldAddEndPuncttrue
\mciteSetBstMidEndSepPunct{\mcitedefaultmidpunct}
{\mcitedefaultendpunct}{\mcitedefaultseppunct}\relax
\EndOfBibitem
\bibitem{TMVA4}
A.~Hoecker {\em et~al.}, \ifthenelse{\boolean{articletitles}}{\emph{{TMVA 4 --- Toolkit for Multivariate Data Analysis with ROOT. Users Guide.}}, }{}\href{http://arxiv.org/abs/physics/0703039}{{\normalfont\ttfamily arXiv:physics/0703039}}\relax
\mciteBstWouldAddEndPuncttrue
\mciteSetBstMidEndSepPunct{\mcitedefaultmidpunct}
{\mcitedefaultendpunct}{\mcitedefaultseppunct}\relax
\EndOfBibitem
\bibitem{Verkerke:2003ir}
W.~Verkerke and D.~P. Kirkby, \ifthenelse{\boolean{articletitles}}{\emph{{The RooFit toolkit for data modeling}}, }{}eConf \textbf{C0303241} (2003) MOLT007, \href{http://arxiv.org/abs/physics/0306116}{{\normalfont\ttfamily arXiv:physics/0306116}}\relax
\mciteBstWouldAddEndPuncttrue
\mciteSetBstMidEndSepPunct{\mcitedefaultmidpunct}
{\mcitedefaultendpunct}{\mcitedefaultseppunct}\relax
\EndOfBibitem
\bibitem{Brun:1997pa}
R.~Brun and F.~Rademakers, \ifthenelse{\boolean{articletitles}}{\emph{{ROOT: An object oriented data analysis framework}}, }{}\href{https://doi.org/10.1016/S0168-9002(97)00048-X}{Nucl.\ Instrum.\ Meth.\  \textbf{A389} (1997) 81}\relax
\mciteBstWouldAddEndPuncttrue
\mciteSetBstMidEndSepPunct{\mcitedefaultmidpunct}
{\mcitedefaultendpunct}{\mcitedefaultseppunct}\relax
\EndOfBibitem
\bibitem{Skwarnicki:1986xj}
T.~Skwarnicki, {\em {A study of the radiative cascade transitions between the Upsilon-prime and Upsilon resonances}}, PhD thesis, Institute of Nuclear Physics, Krakow, 1986, {\href{http://inspirehep.net/record/230779/}{DESY-F31-86-02}}\relax
\mciteBstWouldAddEndPuncttrue
\mciteSetBstMidEndSepPunct{\mcitedefaultmidpunct}
{\mcitedefaultendpunct}{\mcitedefaultseppunct}\relax
\EndOfBibitem
\bibitem{Pivk:2004ty}
M.~Pivk and F.~R. Le~Diberder, \ifthenelse{\boolean{articletitles}}{\emph{{sPlot: A statistical tool to unfold data distributions}}, }{}\href{https://doi.org/10.1016/j.nima.2005.08.106}{Nucl.\ Instrum.\ Meth.\  \textbf{A555} (2005) 356}, \href{http://arxiv.org/abs/physics/0402083}{{\normalfont\ttfamily arXiv:physics/0402083}}\relax
\mciteBstWouldAddEndPuncttrue
\mciteSetBstMidEndSepPunct{\mcitedefaultmidpunct}
{\mcitedefaultendpunct}{\mcitedefaultseppunct}\relax
\EndOfBibitem
\bibitem{Santos:2013gra}
D.~Mart{\'\i}nez~Santos and F.~Dupertuis, \ifthenelse{\boolean{articletitles}}{\emph{{Mass distributions marginalized over per-event errors}}, }{}\href{https://doi.org/10.1016/j.nima.2014.06.081}{Nucl.\ Instrum.\ Meth.\  \textbf{A764} (2014) 150}, \href{http://arxiv.org/abs/1312.5000}{{\normalfont\ttfamily arXiv:1312.5000}}\relax
\mciteBstWouldAddEndPuncttrue
\mciteSetBstMidEndSepPunct{\mcitedefaultmidpunct}
{\mcitedefaultendpunct}{\mcitedefaultseppunct}\relax
\EndOfBibitem
\end{mcitethebibliography}

\newpage
\centerline
{\large\bf LHCb collaboration}
\begin
{flushleft}
\small
R.~Aaij$^{36}$\lhcborcid{0000-0003-0533-1952},
A.S.W.~Abdelmotteleb$^{55}$\lhcborcid{0000-0001-7905-0542},
C.~Abellan~Beteta$^{49}$,
F.~Abudin{\'e}n$^{55}$\lhcborcid{0000-0002-6737-3528},
T.~Ackernley$^{59}$\lhcborcid{0000-0002-5951-3498},
A. A. ~Adefisoye$^{67}$\lhcborcid{0000-0003-2448-1550},
B.~Adeva$^{45}$\lhcborcid{0000-0001-9756-3712},
M.~Adinolfi$^{53}$\lhcborcid{0000-0002-1326-1264},
P.~Adlarson$^{80}$\lhcborcid{0000-0001-6280-3851},
C.~Agapopoulou$^{13}$\lhcborcid{0000-0002-2368-0147}
% ...

{\footnotesize \it

% ...
$^{13}$Universit{\'e} Paris-Saclay, CNRS/IN2P3, IJCLab, Orsay, France\\
% ...
$^{36}$Nikhef National Institute for Subatomic Physics, Amsterdam, Netherlands\\
% ...
$^{45}$Instituto Galego de F{\'\i}sica de Altas Enerx{\'\i}as (IGFAE), Universidade de Santiago de Compostela, Santiago de Compostela, Spain\\
% ...
$^{49}$Physik-Institut, Universit{\"a}t Z{\"u}rich, Z{\"u}rich, Switzerland\\
% ...
$^{53}$H.H. Wills Physics Laboratory, University of Bristol, Bristol, United Kingdom\\
% ...
$^{55}$Department of Physics, University of Warwick, Coventry, United Kingdom\\
% ...
$^{58}$School of Physics and Astronomy, University of Glasgow, Glasgow, United Kingdom\\
$^{59}$Oliver Lodge Laboratory, University of Liverpool, Liverpool, United Kingdom\\
% ...
$^{67}$Syracuse University, Syracuse, NY, United States\\
% ...
$^{80}$Department of Physics and Astronomy, Uppsala University, Uppsala, Sweden, associated to $^{58}$\\
% ...
}
\end{flushleft}

\end{document}